\documentclass[twocolumn,showpacs,preprintnumbers,amsmath,amssymb]{revtex4}
\usepackage{epsfig}

\usepackage{comment}
\usepackage[table]{xcolor}
\usepackage{graphicx}
\usepackage{dcolumn}
\usepackage{bm}
\usepackage{hyperref}
\hypersetup{colorlinks=true,linkcolor=magenta,anchorcolor=green,citecolor=cyan,filecolor=black,menucolor=black,urlcolor=brown}
\usepackage{slashed}

\newcommand{\be}{\begin{equation}}
\newcommand{\ee}{\end{equation}}
\newcommand{\ba}{\begin{eqnarray}}
\newcommand{\ea}{\end{eqnarray}}

\begin{document}

\title{Self-similar solutions for Fuzzy Dark Matter}

\author{Raquel Galazo Garc\'ia}
\affiliation{Universit\'{e} Paris-Saclay, CNRS, CEA, Institut de physique th\'{e}orique, 91191, Gif-sur-Yvette, France}
\author{Philippe Brax}
\affiliation{Universit\'{e} Paris-Saclay, CNRS, CEA, Institut de physique th\'{e}orique, 91191, Gif-sur-Yvette, France}
\author{Patrick Valageas}
\affiliation{Universit\'{e} Paris-Saclay, CNRS, CEA, Institut de physique th\'{e}orique, 91191, Gif-sur-Yvette, France}

\begin{abstract}

Fuzzy Dark Matter (FDM) models admit self-similar solutions which are very different
from the standard Cold Dark Matter (CDM) self-similar solutions and do not converge to the latter
in the semiclassical limit. In contrast with the familiar CDM hierarchical collapse,
they correspond to an inverse-hierarchy blow-up. Constant-mass shells start in the
nonlinear regime, at early times, with small radii and high densities, and expand to reach
at late times the Hubble flow, up to small linear perturbations. Thus, larger masses
become linear first. This blow-up approximately follows the Hubble expansion, so that
the central density contrast remains constant with time, although the width of the
self-similar profile shrinks in comoving coordinates. As in a gravitational cooling
process, matter is ejected from the central peaks through successive clumps.
As in wave systems, the velocities of the geometrical structures and of the
matter do not coincide, and matter slowly moves from one clump to the next,
with intermittent velocity bursts at the transitions.
These features are best observed using the density-velocity representation of the
nonrelativistic scalar field, or the mass-shell trajectories, than with the Husimi
phase-space distribution, where an analogue of the Heisenberg uncertainty principle
blurs the resolution in the position or velocity direction.
These behaviours are due to the quantum pressure and the wavelike
properties of the Schr\"odinger equation. Although the latter has been used as an
alternative to N-body simulations for CDM, these self-similar solutions show that the
semiclassical limit needs to be handled with care.

\end{abstract}

\date{\today}

 \maketitle

\section{Introduction}
\label{sec:introduction}

The lack of experimental evidence in favor of traditional particle-physics explanations for the existence of dark matter (DM), such as the usual WIMP scenario \citep{STRIGARI20131,Schumann_2019}, has led to a resurgence of alternative models. Amongst those, scalar-field models are strong contenders thanks to the host of additional phenomena they could explain or reproduce. One of the most striking is the existence of smooth soliton-like structures of large sizes that could describe the core structure of galactic halos \citep{Schive2014}. 
They could provide a natural explanation for cold-dark-matter (CDM) puzzles such as the core problem, where the small-scale distribution of matter in the CDM scenario appears too peaked compared to observations \citep{Burkert_1995,Block2010}. In this large class of models, whose forefather is certainly the axion originally proposed in relation with the QCD -$\theta$-problem \citep{Peccei1977,Wilczek1978,Weinberg1978},
fuzzy dark matter (FDM) \citep{Hu2000,Hui2017} is  the scalar-field dark-matter model (SFDM) with  the simplest possibility. Indeed in this scenario, the dark matter density at the background level is simply due to a fast-oscillating scalar field in a parabolic potential. On average and therefore on timescales much larger than the fast scalar oscillations, the pressure of the scalar field vanishes.  Other scenarios include additional terms to the potential, such as quartic terms  $V_I(\phi) \propto \lambda \phi^4 $ to describe dark matter self-interactions \cite{Brax:2019fzb}. When gravity and self-interactions are present, the phenomenology of these dark matter models is very rich. Indeed, most of these models admit soliton-like structures whose origin can be easily grasped by comparing the pressure responsible for the various equilibrium configurations
\citep{Goodman:2000tg,Chavanis:2011zi,Chavanis:2011zm,Luu:2018afg}. 
In the self-interaction case, a positive quartic interaction gives rise to a pressure that can lead to equilibria when balancing the gravitational attraction. Similarly in the absence of self-interactions, gravity and the quantum pressure originating from the  gradients of the scalar field can also lead to large-scale equilibria.

Whereas the CDM dynamics are governed by the Vlasov equation, the SFDM dynamics are
governed by the Schr\"odinger equation in the nonrelativistic limit.
This leads to wavelike effects, such as interferences and the Heisenberg principle.
Through the Madelung transformation \cite{Madelung1926}, the Schr\"odinger equation can be mapped to
hydrodynamics in certain regimes \citep{Wallstrom1994} (i.e. outside of regions where the density
vanishes where the mapping becomes ill defined) and these wavelike effects give
rise to a so-called quantum pressure in the Euler equation. As recalled above,
the latter can balance gravity and allows the formation of stable equilibria with
flat density cores, on a size set by the de Broglie wavelength where wavelike effects
are important. On larger scales, wavelike effects are negligible and one recovers
the CDM dynamics. For instance, FDM numerical simulations find virialized halos that
consist of a solitonic flat core embedded in an NFW-like halo \citep{Schive2014,Schwabe2016,Mocz2017,Veltmaat2018}. 
This allows one
to recover the success of the CDM scenario on cosmological scales while having
significant deviations on small scales, which can be of the order of galactic sizes
for a scalar-field mass $m \sim 10^{-22} {\rm eV}$ \citep{Schive2014}.

From a different perspective, because wave effects are expected to be negligible
on scales much larger than the de Broglie wavelength $\lambda_{\rm dB}$,
the Schr\"odinger equation has been used as an alternative to N-body simulations
to compute the CDM dynamics \citep{Widrow1993,Uhlemann2014,Mocz2018,Garny2020}.
Here, one does not wish to investigate FDM but rather
the standard CDM, representing particle dark matter candidates such as  WIMPs, and the 
Schr\"odinger equation is only considered
as a mathematical device to approximate the Vlasov equation. The idea is that
by moving from the six-dimensional phase-space distribution function
$f(\vec r, \vec v)$ to the three-dimensional complex field $\psi(\vec r)$, one may
gain a significant computer time when simulating the evolution of CDM structures. As a result $\lambda_{\rm dB}$ sets a
small-scale resolution cutoff, which is not physical and simply needs to be taken
much smaller than the scale $L$ of interest, like the smoothing scale of a standard
N-body simulation. In other words, one is interested in the semiclassical limit
$\epsilon \to 0$, where $\epsilon \sim \lambda_{\rm dB}/L \propto 1/m$.

Both from the points of view of FDM scenarios and of the Schr\"odinger equation
as a  proxy for  the Vlasov equation, it is interesting to compare
the CDM and FDM dynamics. As recalled above, numerical simulations have
shown that the cosmic webs are similar while virialized halos show a flat
solitonic core inside an NFW envelope. In this paper, we extend the  analytical description of FDM  solutions
 beyond the realm of  static solitons by investigating self-similar solutions.
This allows us to obtain dynamical configurations embedded in the expanding
cosmological background and to compare them with the well-known CDM self-similar
solutions.

In the CDM scenario, the self-similar solutions display a well-known hierarchical
collapse \cite{Fillmore:1984wk,Bertschinger:1985pd,Teyssier1997}. 
Starting at early times with small linear density perturbations on top
of the homogeneous background, with an amplitude that decays as a power law at large
radii, larger mass shells turn around and collapse as time goes on.
For sufficiently steep initial profiles (or stiff equation of state in the
case of a collisional gas), the inner core stabilizes in physical coordinate $r$.
This leads to the  building of a virialized halo, with a power-law density profile in the inner
nonlinear region, that grows in mass and radius with time as more distant shells
separate from the Hubble flow and collapse.
We will find that the FDM self-similar solutions are completely different
and display instead an inverse-hierarchy blow-up.
The system is now nonlinear at early times and becomes linear at late times,
although it does not converge to the usual linear theory; larger masses become linear earlier.
The ejection of matter, due to the wavelike effects associated to the
quantum pressure, is reminiscent of gravitational cooling \citep{Seidel1994,Guzman2006}, 
where energy is evacuated
from the inner core even in the absence of dissipation.
This also implies that the well-known CDM self-similar solutions are not the semiclassical
limit of these FDM self-similar solutions.

The paper is arranged as follows. In Sec.~\ref{sec:eqs-of-motion}, we describe
the dynamics of fuzzy dark matter, its equations of motion and the
semiclassical limit.
In Sec.~\ref{sec:solitons}, we present the static solitons and we derive the
critical exponents that characterize self-similar solutions.
In Sec.~\ref{sec:cosmology}, we study spherical cosmological self-similar solutions,
which converge to the background universe at large radii.
We obtain their large-density asymptotic shape in Sec.~\ref{sec:asymptotic}.
Finally, in Sec.~\ref{conc} we compare with the CDM self-similar solutions,
we discuss the semiclassical limit and conclude.

\section{Equations of motion}
\label{sec:eqs-of-motion}

\subsection{Fuzzy dark matter action}
\label{sec:FDM}

The action of Fuzzy Dark Matter is that of a classical scalar field $\phi$
with minimal coupling to gravity and no self-interactions \citep{Hui2017},
\be
S= \int d^4x \sqrt{-g} \left[ - \frac{1}{2} g^{\mu\nu}\partial_\mu\phi\partial_\nu\phi
- \frac{m^2}{2} \phi^2 \right] .
\ee
In this section and the next one, we focus on astrophysical or galactic
scales where the expansion of the Universe can be neglected and metric fluctuations
are small, so that Newtonian gravity applies. We shall include the Hubble expansion 
in Sec.~\ref{sec:cosmology}. It will turn out that the self-similar exponents are identical
for the Minkowski background and the expanding Eisntein-de Sitter background.
In the nonrelativistic regime, relevant for astrophysical and large-scale structures, 
it is useful to introduce
a complex scalar field $\psi$ by \citep{Hu2000,Hui2017},
\be
\phi = \frac{1}{\sqrt{2m}} \left( \psi \exp^{-imt} + \psi^{*} \exp^{imt} \right).
\ee
This allows us to separate the fast oscillations at frequency $m$ from the slower dynamics
described by $\psi$ that follow the evolution of the density field and of the gravitational potential.
Using $\dot\psi \ll m \psi$ and $\nabla \psi \ll m \psi$ and neglecting the expansion of the universe, the Klein-Gordon equation for $\phi$
leads to the Schr\"odinger equation for $\psi$,
\be
i \frac{\partial\psi}{\partial t} = -\frac{1}{2 m} \nabla^2 \psi + m \Phi_{\rm N} \, \psi,
\label{eq:Schrodinger-real-1}
\ee
where $\Phi_{\rm N}$ is the gravitational potential, given by the Poisson equation
\be
\nabla^2 \Phi_{\rm N} = 4\pi {\cal G}_{\rm N} \rho, \;\;\; \rho = m \psi \psi^* ,
\label{eq:Poisson-real-1}
\ee
 where $\rho$ is the FDM density.
 
 An interesting feature of this Schr\"odinger-Poisson (SP) system is that it is invariant
 under the scaling law \citep{Guzman:2003kt} 
 \be
\left\lbrace t, \vec r, \Phi_{\rm N}, \psi , \rho \right\rbrace \rightarrow \left\lbrace \lambda^{-2}t, \lambda^{-1} \vec r , \lambda^{2}\Phi_{\rm N}, \lambda^{2}\psi , \lambda^4 \rho \right\rbrace .
\label{eq:res1}
\ee
This means that once we have obtained an equilibrium or a dynamical solution,
a complete family of solutions can be derived through this scaling transformation.

\subsection{Semiclassical limit}
\label{sec:semiclassical}

It is often convenient to use dimensionless coordinates suited to the system under study.
Having in mind galactic cores or astrophysical objects, such as FDM halos, let us consider
a system of typical length $L$, timescale $T$ and velocity $V=L/T$.
For systems dominated by gravity and not far from equilibrium,
the virial theorem implies that the gravitational potential is of the order of $V^2$.
Then, using the rescalings
\ba
&& t = T \, \tilde t , \;\; \vec r= L \, \tilde{ \vec r} , \;\; \rho = \frac{\tilde\rho}{{\cal G}_{\rm N} T^2} ,
\;\; \Phi_{\rm N} = (L/T)^2 \, \tilde\Phi_{\rm N} ,  \nonumber \\
&& \psi = \frac{\tilde\psi}{\sqrt{{\cal G}_{\rm N} m} T} ,
\label{eq:rscaling-eps}
\ea
we obtain for the dimensionless variables denoted with a tilde sign the Schr\"odinger-Poisson (SP) system
\ba
&& i \epsilon \frac{\partial\tilde\psi}{\partial\tilde t} = -\frac{\epsilon^2}{2} \tilde\nabla^2 \tilde\psi
+ \tilde\Phi_{\rm N} \, \tilde\psi ,
\label{eq:Schrod-eps}  \\
&&  \tilde\nabla^2 \tilde\Phi_{\rm N} = 4 \pi \tilde\rho , \;\;\; \tilde\rho= \tilde\psi \tilde\psi^* ,
\label{eq:Poisson-eps}
\ea
 where $\epsilon$ is given by
\be
\epsilon = \frac{T}{m L^2} .
\ee
Comparing with the de Broglie wavelength $\lambda_{\rm dB} = 2\pi/(m V)$,
we have
\be
\epsilon \sim \frac{\lambda_{\rm dB}}{L} .
\ee
Thus, $\epsilon$, which appears in the dimensionless Schr\"odinger equation
(\ref{eq:Schrod-eps}) as $\hbar$ in quantum mechanics, corresponds to the ratio between
the de Broglie wavelength $\lambda_{\rm dB}$ and the size of the system.
FDM can alleviate tensions of the $\Lambda$CDM scenario with observations on galactic scales
by building DM cores (solitons) of radius $\lambda_{\rm dB}$, where the wavelike behavior
is important. On larger scales the system behaves like a collection of particles and
numerical simulations show that the core is embedded in an NFW-like halo, as in
the standard CDM scenario \citep{Schive2014,Schwabe2016,Mocz2017,Veltmaat2018}.
Requiring that $\lambda_{\rm dB}$ should be of the order of the galactic scales sets the scalar
field mass to be of the order of $10^{-22} \, {\rm eV}$ \citep{Schive2014}.
This corresponds to $\epsilon \sim 1$, as the wavelike effects are important on the scales of
interest.

In the following, we work with the dimensionless variables (\ref{eq:rscaling-eps})
and we remove the tildes for simplicity of notation.

The ``semiclassical'' limit $\epsilon \to 0$ corresponds to the case
where FDM behaves like DM on all relevant scales. In this regime, the SP system
(\ref{eq:Schrod-eps})-(\ref{eq:Poisson-eps}) has been advocated as an alternative
framework for cosmological simulations of CDM, instead of the standard N-body simulations
that aim at reproducing the Vlasov equation \citep{Widrow1993,Uhlemann2014,Mocz2018,Garny2020}.
Indeed, in the semiclassical limit $\epsilon \to 0$
we expect mean quantities, such as the density averaged over the fast oscillations at
frequency $m$, to converge to those that would be obtained from the Vlasov equation.
More precisely, defining the Wigner distribution \citep{Wigner1932} by
\be
f_{\rm W}(\vec r, \vec v) = \int \frac{d\vec r \,'}{(2\pi)^3} \, e^{i \vec v \cdot \vec r \,' }
\psi\left(\vec r - \frac{\epsilon}{2} \vec r\,' \right)
\psi^*\left(\vec r + \frac{\epsilon}{2} \vec r\,' \right)  ,
\label{eq:fW-def}
\ee
and its coarse-grained average, the Husimi distribution \citep{Husimi1940}, by
\ba
f_{\rm H}(\vec r, \vec v) & = & \int \frac{d\vec r\,' d\vec v\,'}{(2\pi\epsilon)^3\sigma_r^3 \sigma_v^3} \,
e^{-(\vec r - \vec r\,')^2/(2\epsilon\sigma_r^2) - (\vec v - \vec v\,')^2/(2\epsilon\sigma_v^2)}
\nonumber \\
&& \times f_{\rm W}(\vec r\,',\vec v\,') ,
\label{eq:fH-def}
\ea
one can check that both distributions obey an equation of motion that differs from
the Vlasov equation by terms of order $\epsilon$ and higher
\citep{Skodje1989,Widrow1993,Uhlemann2014}.
For $\sigma_r \to 0$ and $\sigma_v \to 0$ the Husimi distribution
converges to the Wigner distribution, which can take negative values and display
fast oscillations, in contrast with classical phase-space distributions.
However, for $\sigma_r \sigma_v \geq 1/2$ the Husimi distribution is positive
\citep{Cartwright1976},
which provides a better comparison with classical physics.
The lower bound $\sigma_r \sigma_v \geq 1/2$ is related to the Heisenberg
uncertainty principle. It implies that we cannot obtain a classical description
with arbitrarily high accuracy simultaneously on both space and momentum
coordinates. In the following, we shall take $\sigma_r \sigma_v = 1/2$ for numerical
computations, as this is the best available choice. Then, as will be seen in 
Figs.~\ref{fig:husimi-10-100} and \ref{fig:husimi-0p99} below, a smaller $\sigma_r$
will give a better spatial accuracy but a worse velocity resolution.
For $\sigma_r \sigma_v =1/2$, the Husimi distribution (\ref{eq:fH-def}) also reads as 
\be
f_{\rm H}(\vec r, \vec v) = \left| \int \frac{d\vec r\,'}{(2\pi\epsilon)^{9/4} \sigma_r^{3/2}} \,
e^{-(\vec r - \vec r\,')^2/(4\epsilon\sigma_r^2) - i \vec v \cdot \vec r\,' / \epsilon} \psi(\vec r\,')
\right|^2
\label{eq:fH-positive}
\ee
which shows that it is always positive, like classical phase-space distributions.
This is why the semiclassical limit is better analysed in terms of the Husimi distribution
than in terms of the Wigner distribution, which is not definite positive and typically shows fast
oscillations.
However, the semiclassical limit remains a subtle problem 
\citep{Jin2011} and only coarse-grained quantities,
averaged over the fast oscillations, are expected to converge to their classical counterparts,
while the wave function $\psi$ keeps strong oscillations on the increasingly small wavelength
$\lambda_{\rm dB} \sim \epsilon L$ as $\epsilon\to 0$, at fixed macroscopic scale $L$.

\subsection{Hydrodynamical picture}
\label{sec:hydro}

Taking the Madelung transformation \citep{Madelung1926}, $\psi \to \{\rho,S, \vec v\}$,
\be
\psi = \sqrt{\rho} \, e^{i S/\epsilon} , \;\;\; \vec v = \nabla S ,
\label{eq:Madelung}
\ee
where the amplitude $\rho$ plays the role of the scalar density and $\vec v$ that of the scalar
velocity, the real and imaginary parts of the Schr\"odinger equation lead to the continuity and
Hamilton-Jacobi equations,
 \ba
&& \frac{\partial\rho}{\partial t} + \nabla \cdot ( \rho \nabla S) = 0 ,
\label{eq:continuity-S} \\
&&  \frac{\partial S}{\partial t} + \frac{1}{2} (\nabla S)^2 = - ( \Phi_{\rm N}+\Phi_{\rm Q} ),
\label{eq:Hamilton-Jacobi}
\ea
where we have introduced the so-called ``quantum pressure'' \citep{Spiegel1980,Chavanis:2011zi,Marsh2015} $\Phi_{\rm Q}$, given by
\be
\Phi_{\rm Q}=  - \frac{\epsilon^2}{2} \frac{\nabla^2\sqrt{\rho}}{\sqrt{\rho}} .
\label{eq:quantum-pressure}
\ee
In terms of the curl-free velocity field $\vec v$, this gives the hydrodynamical continuity and
Euler equations,
\ba
&& \frac{\partial \rho}{\partial t} + \nabla \cdot (\rho\vec{v}) = 0 ,
\label{eq:continuity-v} \\
&& \frac{\partial \vec{v}}{\partial t} + (\vec{v} \cdot \nabla) \vec{v} = - \nabla
\left( \Phi_{\rm N}  + \Phi_{\rm Q} \right)  .
\label{eq:euler-1}
\ea
The Poisson equation still reads
\be
\nabla^2 \Phi_{\rm N} = 4\pi \rho .
\label{eq:poisson-2}
\ee

We can see that the parameter $\epsilon$, which sets the relevance of wavelike effects,
only appears as the prefactor of the quantum pressure (\ref{eq:quantum-pressure}).
Thus, in the limit $\epsilon \to 0$, we recover the usual continuity and Euler equations,
which also describe CDM on large scales where shell crossing can be neglected.
This is another manifestation of the fact that in the semiclassical limit, $\epsilon \to 0$,
or on large scales where the quantum pressure is suppressed by the Laplacian $\nabla^2$,
SFDM behaves like CDM.
However, the limit $\epsilon \to 0$ is not uniform and can break down on small scales:
as $\epsilon \to 0$ the fields can show gradients that become increasingly steep,
i.e. they vary on increasingly small scales or order $\epsilon$.
This counterbalances the $\epsilon^2$ prefactor in Eq.(\ref{eq:quantum-pressure})
and the quantum pressure cannot be neglected everywhere in space.

As pointed out in \cite{Wallstrom1994} for instance, the hydrodynamical equations 
(\ref{eq:continuity-v})-(\ref{eq:euler-1}) are not strictly equivalent
to the Schr\"odinger equation (\ref{eq:Schrod-eps}), because the mapping
$\psi \leftrightarrow \{\rho,S, \vec v\}$ becomes ill-defined when the density vanishes
(the phase is no longer unique). This can lead to the generation of vorticity along the
lines $\rho=0$, whereas the velocity field $\vec v$ defined in Eq.(\ref{eq:Madelung})
is always curl-free, being the gradient of a scalar.
In such cases, one can no longer work with the hydrodynamical variables $\{\rho,\vec v\}$
and one must go back to the wave function $\psi$.
Nevertheless, in regimes where the density does not vanish (or when such discrepancies
can be neglected) the hydrodynamical picture remains useful, as it is simpler to interpret
and provides a more direct comparison with the density and velocity fields
used to describe the cosmological distribution of DM.
In this paper, we shall focus on spherically symmetric solutions, where the equivalence is exact.
Indeed, this makes the problem one-dimensional in space and any radial velocity $v_r$
can be written as the gradient of a phase $S$, so that one can go back from the hydrodynamical
picture to the Schr\"odinger picture with $S = \int dr \, v_r$.  
Besides, we shall find that for our self-similar solutions the density never vanishes if $\epsilon>0$.

\subsection{Convergence to the classical distribution and multistreaming}
\label{sec:convergence-semiclassical}

As FDM has been used as an alternative tool from N-body simulations to study CDM,
and because we shall find that the properties of FDM self-similar solutions are quite different
from those of CDM self-similar solutions, we discuss here in more details the semiclassical
limit. In particular, we present the link between the wave function $\psi$, the Husimi phase
space distribution and its semiclassical limit, and the hydrodynamical picture.
From Eqs.(\ref{eq:fH-positive}) and (\ref{eq:Madelung}), we write
\be
f_{\rm H}(\vec r, \vec v) = \left| \int \!\! \frac{d\vec r\,'}{(2\pi \epsilon)^{9/4} \sigma_r^{3/2}} 
\sqrt{ \rho (\vec r\,') } \, e^{-\frac{(\vec r\,'-\vec r)^2}{4\sigma^2_r \epsilon} 
- \frac{i}{\epsilon} \vec v \cdot \vec r\,' + \frac{i}{\epsilon} S(\vec r\,') } \right|^2
\ee
Making the change of variables $\vec r\,' = \vec r + \sqrt{\epsilon} \vec r\,''$ and
expanding $\rho(\vec r\,')$ and $S(\vec r\,')$, we obtain at leading order over $\epsilon$
\be
f_{\rm H}  \simeq  \rho (\vec r) \left| \int \!\! \frac{d\vec r\,''}{(2\pi)^{9/4} \epsilon^{3/4}\sigma_r^{3/2}} 
e^{\frac{i}{\sqrt{\epsilon}} (S_{,j} - v_j) r_j'' - \frac{r''^2}{4\sigma_r^2}+\frac{i}{2} S_{,jk} r_j'' r_k''}
\right|^2  
\ee
where we sum over the spatial indices $j,k$ and we denoted the spatial derivatives
$S_{,j} = \partial_j S$ and $S_{,jk} = \partial_j\partial_k S$.
Being real symmetric, the matrix $S_{,jk}$ is diagonalizable with real eigenvalues $s_j$.
Then, the complex matrix $M$ defined by
\be
M_{jk} = \delta_{jk} - i 2 \sigma_r^2 S_{,jk} 
\label{eq:Mij-def}
\ee
is also diagonalizable with eigenvalues $m_j = 1 - i 2 \sigma_r^2 s_j$,
with strictly positive real parts. Therefore, we can perform the Gaussian integral,
\ba
f_{\rm H}(\vec r, \vec v) & \simeq & \rho(\vec r) \left( \frac{2\sigma_r^2}{\pi\epsilon}\right)^{3/2}
\left( \det(M) \det(M^*) \right)^{-1/2} \nonumber \\
&& \times e^{- \frac{\sigma_r^2}{\epsilon} ( M^{-1} + M^{*-1} )_{jk}
(S_{,j}-v_j) (S_{,k}-v_k) } .
\ea
Using (\ref{eq:Mij-def}) we also have
\be
M^{-1} + M^{*-1} = M^{-1} ( M + M^* ) M^{*-1} = 2 M^{-1} M^{*-1} .
\ee
This is then a diagonalizable matrix with strictly positive real eigenvalues.
Therefore, in the limit $\epsilon \to 0$ the Gaussian velocity factor gives a Dirac term
with a normalization that cancels the determinant prefactor,
\be
\epsilon \to 0 : \;\;\; f_{\rm H}(\vec r, \vec v) = \rho(\vec r) \, \delta_D(\vec v - \nabla S) .
\label{eq:fH-semi-classical-Dirac}
\ee
This is the classical distribution in phase space for a single-stream flow with density $\rho$
and velocity $\vec v = \nabla S$.
Of course, this result only holds if the wave function only varies on the macroscopic scale
of interest $L$ and does not show structures at scale $\sqrt{\epsilon}L$, so that we can use 
a Taylor expansion for the density and the phase.
This is violated if the dynamics generate structures on increasingly small scales as
$\epsilon \to 0$. This is why the semiclassical limit can be a delicate matter.

In the multistreaming regime, the wave function reads as \citep{Jin2011}
\be
\psi = \sum_{{\rm stream} \; j} \sqrt{\rho_j(\vec r)} \, e^{i S_j(\vec r)/\epsilon} ,
\ee
where we sum over the streams, and we obtain
\be
\epsilon \to 0 : \;\;\;  f_{\rm H}(\vec r, \vec v) = \sum_j \rho_j(\vec r) \, \delta_D(\vec v - \nabla S_j) .
\ee
Indeed, the cross-terms that arise from the modulus squared show two Gaussian velocity
factors with well separated peaks and are therefore negligible in the limit $\epsilon\to 0$.
At least locally we can always write $\psi$ in the form (\ref{eq:Madelung}). It is interesting
to see why the derivation (\ref{eq:fH-semi-classical-Dirac}) fails in this case.
Let us choose for simplicity two streams in a one-dimensional system, 
with constant densities $\rho_1$ and $\rho_2$, constant velocities $v_1=0$ and $v_2$,
and hence $S_1=0$ and $S_2=v_2 x$,
\be
\psi = \sqrt{\rho_1} + \sqrt{\rho_2} e^{i v_2 x/\epsilon} = \sqrt{\rho} \, e^{i S/\epsilon} .
\ee
This gives for the total density $\rho$ and phase $S$ the expressions
\ba
&& \rho = \rho_1 + \rho_2 + \sqrt{\rho_1\rho_2} 2 \cos(v_2 x/\epsilon) , \nonumber \\
&& S = \epsilon \arccos \left[ \frac{1}{\sqrt{\rho}} \left( \sqrt{\rho_1} 
+ \sqrt{\rho_2} \cos (v_2 x/\epsilon) \right) \right] .\;\;\;
\ea
We can see that the total density $\rho$ and phase $S$ now show fast oscillations at scale
$\epsilon$. 
Therefore, the Gaussian approximation used above no longer applies
if $\psi$ is written as $ \sqrt{\rho} e^{i S/\epsilon}$. It however applies on each of the two terms
of the expression $\sqrt{\rho_1} + \sqrt{\rho_2} e^{i v_2 x/\epsilon}$, as their densities 
$\rho_j$ and phases $S_j$ do not show fast oscillations.

This illustrates that, while by going from the 6D phase space of the classical distribution
$f(\vec r,\vec v)$ to the 3D configuration space of the wave function $\psi(\vec r)$,
we can hope to obtain a competitive tool to simulate CDM as compared with usual
N-body simulations, the difficulties associated with shell crossings reappear as small-scale
oscillations. Thus, one needs a high accuracy to resolve the different streams and this also
sets a practical lower bound on the semiclassical parameter $\epsilon$ to avoid
too large computer times. 

In the self-similar solutions studied in this paper, we do not have multistreaming but the
width of the solutions shrinks with $\epsilon$, as shown by the scaling (\ref{eq:eta-def-x}) below.
This again implies that the semiclassical limit is not given by a Gaussian approximation as above
and it is not trivial. This is why we shall not recover the CDM self-similar solutions in the limit
$\epsilon\to 0$. Instead, these solutions vanish as their width becomes infinitesimal, while
never reaching a classical regime.

\section{Equilibrium and self-similar solutions}
\label{sec:solitons}

\subsection{Static equilibria: solitons}
\label{sec:soliton}

Static equilibrium profiles, called solitons, have a zero velocity $\vec v$.
This leads through the Euler equation (\ref {eq:euler-1}) to the hydrostatic equilibrium
condition \cite{Chavanis:2011zi}
\be
\Phi_{\rm N} + \Phi_{\rm Q} = \alpha ,
\label{eq:hydrostatic}
\ee
where $\alpha$ is a constant.
Thus, the soliton arises from the balance between the repulsive quantum pressure and
the attractive force of gravity. Then, the Hamilton-Jacobi equation (\ref{eq:Hamilton-Jacobi})
leads to
\be
S= - \alpha t , \;\; \mbox{hence} \;\; \psi = e^{-i \alpha t/\epsilon} \psi_{\rm sol}(r) , \;\;
\rho_{\rm sol} = \psi_{\rm sol}^2 ,
\ee
where we look for spherically symmetric solutions.
Substituting into the Schr\"odinger equation (\ref{eq:Schrod-eps}) gives
\be
\epsilon^2 \nabla^2 \psi_{\rm sol} = 2 ( \Phi_{\rm N} - \alpha) \psi_{\rm sol} .
\label{eq:soliton-eq}
\ee
Coupled with the Poisson equation, $\nabla^2 \Phi_{\rm N} = 4\pi \psi_{\rm sol}^2$,
this nonlinear system determines the soliton profile, with the boundary conditions
$\psi_{\rm sol}'=0$ at $r=0$ and $\psi_{\rm sol} \to 0 $ for $r \to \infty$.
Different values of $\alpha$ correspond to different values of the soliton mass and central
density.
These different profiles are related through the scaling law (\ref{eq:res1}).

\subsection{Self-similar exponents}
\label{sec:self-similar}

As for CDM, in order to go beyond static equilibrium profiles, an interesting approach is to
look for self-similar solutions. This provides time-dependent solutions that can be well
understood with semi-analytical tools.

\subsubsection{Field picture}
\label{subsubsec:field}

In terms of the complex field $\psi$, and of the gravitational potential $\Phi_{\rm N}$,
we look for solutions of the form
\be
\psi = t^{-\alpha} f \left( \frac{r}{t^{\beta}} \right), \qquad
\Phi_{\rm N} = t^{-\mu} h \left( \frac{r}{t^{\beta}} \right) ,
\label{eq:field-self-similar}
\ee
where $f$ and $h$ are unknown functions to be determined, as well as the scaling exponents
$\alpha$, $\beta$ and $\mu$.
Substituting into the SP system (\ref{eq:Schrod-eps})-(\ref{eq:Poisson-eps}),
we obtain
\ba
- i \epsilon t^{-\alpha-1} ( \alpha f + \beta \eta f' ) & = & - \frac{\epsilon^2}{2}
t^{-\alpha -2\beta} \left( f'' + \frac{2}{\eta} f' \right) \nonumber \\
&& + t^{-\mu-\alpha} h f , \label{eq:psi-f-1}
\ea
\be
t^{-\mu-2\beta} \left( h'' + \frac{2}{\eta} h' \right) = 4\pi t^{-\alpha-\alpha^*} f f^* ,
\label{eq:psi-h-1}
\ee
where the prime denotes the derivative with respect to the new variable $\eta = r/t^{\beta}$.
The compatibility conditions for these equations to be expressed in terms of $\eta$ only
give the values of the scaling exponents,
\be
\beta= 1/2 , \;\; \mu= 1, \;\; \alpha = 1 + i \, b
\label{eq:psi-exponents}
\ee
where $b$ is a real undetermined parameter. Thus, the fields take the form
\be
\psi = t^{-1-i b} f \left( \frac{r}{\sqrt{t}} \right) , \qquad
\Phi_{\rm N} = t^{-1} h \left( \frac{r}{\sqrt{t}} \right) .
\label{eq:psi-self}
\ee
We recognize the diffusive scaling $\sqrt{t}$ due to the Laplacian in the Schr\"odinger equation.
The functions $f$ and $h$ must then be determined by solving the ordinary differential equations 
(\ref{eq:psi-f-1})-(\ref{eq:psi-h-1}).

 \subsubsection{Fluid picture}

We can check that we recover the same self-similar exponents by working in the hydrodynamical
picture. Looking for solutions of the form
\be
\rho = t^{-\alpha} f\left(\frac{r}{t^{\beta}}\right), \hspace{0.5em}
v = t^{-\delta}g\left(\frac{r}{t^{\beta}}\right),  \hspace{0.5em}
\Phi_{\rm N} =t^{-\mu}h\left(\frac{r}{t^{\beta}}\right) ,
\label{eq:fluid-self-similar}
\ee
and substituting into the continuity, Euler and Poisson equations
(\ref{eq:continuity-v})-(\ref{eq:poisson-2}), we obtain

\be
- t^{-\alpha-1} \left( \alpha f + \beta\eta f' \right) + t^{-\alpha-\beta-\delta} \left(
\frac{2}{\eta}fg+f'g+fg' \right) =0 ,
\ee
\ba
&& -t^{-\delta-1} \left( \delta g + \beta\eta g' \right) + t^{-2\delta-\beta} gg' = -t^{-\mu-\delta} h'
\nonumber \\
&& + \frac{\epsilon^2}{4} t^{-3\beta} \frac{d}{d\eta} \left( \frac{f''}{f}+\frac{2}{\eta}\frac{f'}{f}
-\frac{1}{2}\left(\frac{f'}{f}\right)^2 \right) ,
\ea
\be
t^{-\mu-2\beta}  \left( h'' + \frac{2}{\eta}h' \right) = 4\pi t^{-\alpha} f .
\ee
The compatibility conditions now give
\be
\beta = 1/2 , \;\; \mu = 1, \;\; \alpha = 2, \;\; \delta= 1/2 ,
\ee
and the fields take the form
\be
\rho = t^{-2} f \left( \frac{r}{\sqrt{t}} \right), \;\;
v =  t^{-1/2} g \left( \frac{r}{\sqrt{t}} \right) , \;\;
\Phi_{\rm N} = t^{-1} h \left( \frac{r}{\sqrt{t}} \right) .
\label{eq:hydro-self}
\ee
Using the relation $\vec v = \vec \nabla S$ and the Hamilton-Jacobi equation (\ref{eq:Hamilton-Jacobi}),
we obtain the phase $S$ from the velocity $\vec v$ as
\be
S = s \left( \frac{r}{\sqrt{t}} \right) + c_0 \ln t + c_1 , \;\;\; \mbox{with} \;\;\; s' = g
\ee
and $c_0$ and $c_1$ are undetermined real constants.
This gives for the complex field $\psi$, using the Madelung transformation
(\ref{eq:Madelung}),
\be
\psi = t^{-1} \sqrt{ f\left( \frac{r}{\sqrt{t}} \right) } e^{ i [ s(r/\sqrt{t}) + c_0 \ln t + c_1 ] /\epsilon} ,
\ee
which takes the same scaling form as the previous result (\ref{eq:psi-self})
(with different meanings for the function $f$ and the parameters $b,c_0$ and $c_1$).

\section{Cosmological self-similar solutions}
\label{sec:cosmology}

\subsection{Cosmological background}

As we are interested in SFDM in the cosmological context, we now focus
on self-similar solutions within a cosmological framework.
As for CDM, such solutions can only be found in cosmological eras where the
scale factor is a power law of time, so that no specific time or length scale is introduced
by the cosmological background, which would break the self-similarity.
Therefore, as for CDM, we focus on the case of the Einstein-de Sitter universe,
which applies to the matter era when most large-scale structures are formed,
until $z \sim 1$.
On scales much smaller than the horizon, the dynamics can be described by Newtonian
gravity. The scale factor grows as $a \propto t^{2/3}$ and the Hubble expansion reads
as
\be
H = \frac{2}{3 t} , \;\; a = t^{2/3} ,
\ee
in dimensionless units.
Then, the background density $\bar \rho$, the Hubble-flow radial velocity $\bar v$ and the
background Newtonian potential $\bar\Phi_{\rm N}$ read
\be
\bar\rho = \frac{1}{6\pi  t^{2}} , \;\;  \bar v= \frac{2r}{3t} , \;\;
\bar\Phi_{\rm N} = \frac{r^2}{9 t^2} .
\ee
We can check that these expressions are solutions of the continuity, Euler and Poisson
equations (\ref{eq:continuity-v})-(\ref{eq:poisson-2}).
These background expressions also take the self-similar forms (\ref{eq:hydro-self}).
This means that we can also look for self-similar solutions associated with perturbations
around this expanding background.

\subsection{Comoving coordinates}
\label{sec:comoving}

For convenience, we work in the hydrodynamical picture as it facilitates the comparison with
the standard CDM scenario. As usual, we introduce the comoving spatial coordinates
$\vec x = \vec r / a$, and we write the density and velocity fields and the gravitational potential as
\be
\rho = \bar \rho (1+\delta) , \;\;
\vec v = \bar{\vec v} + \vec u, \;\;  \Phi_{\rm N} = \bar\Phi_{\rm N}+\varphi_{\rm N} ,
\label{eq:fields-perturbation}
\ee
where $\delta$ is the density contrast and $\vec u$ the peculiar velocity.

Substituting into the continuity, Euler and Poisson equations
(\ref{eq:continuity-v})-(\ref{eq:poisson-2}), we obtain the usual comoving fluid equations
\ba
&& \frac{\partial \delta }{\partial t} + \frac{1}{a}\nabla_x \cdot \left[(1+\delta)\vec{u} \right]=0,
\label{eq:continuity-2} \\
&& \frac{\partial \vec{u} }{\partial t} + \frac{1}{a} (\vec u \cdot {\nabla}_x) \vec{u} + H\vec{u}
= - \frac{1}{a}{\nabla}_x(\varphi_{\rm N} + \Phi_{\rm Q}) , \;\;\;
\label{eq:euler-2} \\
&& \nabla^2_{x} \varphi_{\rm N} = \frac{2}{3}\frac{\delta}{a} ,
\label{eq:poisson-3}
\ea
except for the additional quantum pressure on the right-hand side of the Euler equation,
which is given by
\be
\Phi_{\rm Q} = - \frac{\epsilon^2}{2 a^2} \frac{\nabla_x^2 \sqrt{\rho}}{\sqrt{\rho}} .
\label{eq:Phi-Q-rho}
\ee

These hydrodynamical equations can also be obtained more rigorously from the action
of the scalar field $\phi$, written in the expanding metric with linear-gravity perturbations
around the FLRW metric \citep{Brax:2019fzb}. 
In the nonrelativistic limit, the comoving Schr\"odinger equation then becomes
\be
i\epsilon \frac{\partial\psi}{\partial t} = - \frac{\epsilon^2}{2 a^2} \nabla_x^2 \psi + \varphi_{\rm N} \psi ,
\label{eq:psi-eq-comoving}
\ee
where we have factored out the term $1/(\sqrt{6\pi} t)$ from the amplitude of $\psi$,
associated with the decrease of the background density \citep{Widrow1993}.

This comoving Schr\"odinger-Poisson system describes the dynamics 
of large-scale structures well inside the Hubble radius, where relativistic corrections
can be neglected. This is a standard approximation in studies of the formation
of large-scale structures in the matter era
\citep{1980lssu.book.Peebles,peacock_1998,mo_van_den_bosch_white_2010}, 
both in analytical works and N-body simulations.
This is because the scale where the density field becomes nonlinear
($\lesssim 10$ Mpc) is much smaller than the Hubble horizon.
As usual, these Newtonian equations of motion are then extended to infinite
space (e.g. continuous Fourier transforms involve an integral over all space; 
alternatively, N-body simulations take a finite box with periodic boundary conditions.)
This procedure is valid as long as one focuses on scales much smaller than the
Hubble radius.

Writing the comoving Madelung transformation as
\be
\psi = \sqrt{1+\delta} \; e^{iS/\epsilon} , \;\;\; \vec p = a \vec u = \vec \nabla_x S ,
\ee
where $\vec p$ is the comoving momentum and $\vec u$ the peculiar velocity,
we recover the continuity and Euler equations (\ref{eq:continuity-2})-(\ref{eq:euler-2}),
while the comoving Hamilton-Jacobi equation for the phase $S$ reads
\be
\frac{\partial S}{\partial t} + \frac{(\vec \nabla_x S)^2}{2 a^2} = - \varphi_{\rm N} - \Phi_{\rm Q} .
\label{eq:Hamilton-Jacobi-x}
\ee

The comoving Wigner distribution now becomes
\be
f_{\rm W}(\vec x, \vec p) = \int \frac{d\vec x\,'}{(2\pi)^3} \, e^{i \vec p \cdot \vec x \,' }
\psi\left(\vec x - \frac{\epsilon}{2} \vec x\,' \right)
\psi^*\left(\vec x + \frac{\epsilon}{2} \vec x\,' \right)  .
\ee
Up to corrections of order $\epsilon$, it satisfies the comoving Vlasov equation,
\be
\frac{\partial f_{\rm W}}{\partial t} + \frac{\vec p}{a^2} \cdot \frac{\partial f_{\rm W}}{\partial\vec x}
- \vec \nabla_x\varphi_{\rm N} \cdot \frac{\partial f_{\rm W}}{\partial\vec p} + {\cal O}(\epsilon) = 0 .
\label{eq:Vlasov-Wigner}
\ee
The background comoving field $\bar\psi$ and distribution $\bar f_{\rm W}$ are
\be
\bar\psi = 1 , \;\;\; \bar S= 0, \;\;\; \bar f_{\rm W} = \delta_D(\vec p) ,
\ee
which coincides with the phase-space distribution of the CDM background.

\subsection{Self-similar coordinates}
\label{sec:self-coordinates}

In agreement with (\ref{eq:hydro-self}), we can check that spherical self-similar solutions
will be of the form
\ba
&& \delta(x,t) = \hat\delta(\eta) , \;\; u(x,t) = \epsilon^{1/2} t^{-1/2} \, \hat u(\eta) , \nonumber \\
&& \varphi_{\rm N}(x,t) = \epsilon t^{-1} \, \hat\varphi_{\rm N}(\eta) , \;\;
\Phi_{\rm Q}(x,t) = \epsilon t^{-1} \, \hat\Phi_{\rm Q}(\eta) , \nonumber \\
&& \delta M(x,t) = \epsilon^{3/2} t^{-1/2} \, \delta\hat M(\eta) ,
\label{eq:self-r-x}
\ea
where we defined the perturbed mass $\delta M$ by
\be
\delta M(r) = 4\pi \int_0^r dr \, r^2 \delta\rho(r) = \frac{2}{3} \int_0^x dx \, x^2 \delta(x) ,
\ee
and we introduced the scaling variable
\be
\eta = \frac{t^{1/6} x}{\epsilon^{1/2}} = \frac{r}{\sqrt{\epsilon t}} .
\label{eq:eta-def-x}
\ee
This scaling is consistent with the self-similar exponents obtained in
Sec.~\ref{sec:self-similar} and  we included in addition the scaling in $\epsilon$.
Therefore, the self-similar exponents are identical for the Minkowski
and Einstein-de Sitter backgrounds.

Thus, the characteristic length scale grows as $\sqrt{t}$ in physical units but decreases
as $t^{-1/6}$ in comoving units.
The associated mass decreases as $M \propto 1/\sqrt{t}$.
Therefore, these self-similar solutions will be very different from the ones obtained
for CDM. Their size grows in physical units but more slowly than the scale factor,
so that they actually shrink in comoving units. Thus, their mass also decreases with time,
whereas CDM self-similar solutions grow both in comoving size and in mass
\citep{Fillmore:1984wk,Bertschinger:1985pd}.
This is due to the diffusive scaling $r \sim \sqrt{t}$, which does not depend on the
shape of the self-similar solution nor on the Einstein-de Sitter expansion.

In terms of the scaling variable $\eta$, the Euler equation reads
\be
\frac{1}{6} ( \hat u + \eta \hat u' ) + \hat u \hat u' + \hat\varphi_{\rm N}' + \hat\Phi_{\rm Q}' = 0 ,
\label{eq:Euler-eta}
\ee
which can be integrated as
\be
\frac{1}{6} \eta \hat u + \frac{1}{2} \hat u^2 + \hat\varphi_{\rm N} + \hat\Phi_{\rm Q} = 0 ,
\label{eq:Bernoulli}
\ee
where we used the boundary condition that all fields vanish at infinity, where there is convergence
to the cosmological background.
The fact that the Euler equation can be integrated to give this Bernoulli-like equation
is related to the fact that the Euler equation itself actually derived from the Hamilton-Jacobi
equation (\ref{eq:Hamilton-Jacobi}).
As compared with the hydrostatic equilibrium (\ref{eq:hydrostatic}), which determined the
soliton profiles from the balance between gravity and quantum pressure,
the Bernoulli equation (\ref{eq:Bernoulli}) includes the impact of the kinetic energy,
as we consider dynamical solutions.

In terms of the $\psi$ field, we obtain the self-similar scalings
\be
S = \epsilon \hat S(\eta) , \;\;\; \psi = \hat\psi(\eta) = \sqrt{1+\hat\delta} \; e^{i\hat S} .
\ee
Comparing the self-similar form of the Hamilton-Jacobi equation (\ref{eq:Hamilton-Jacobi-x})
with the Bernoulli equation (\ref{eq:Bernoulli}) gives
\be
\hat S ' = \hat u .
\ee

Defining as in (\ref{eq:eta-def-x}) and (\ref{eq:self-r-x}) the rescaled position $\vec\eta$
and momentum $\vec\nu$,
\be
\vec x = \epsilon^{1/2} t^{-1/6} \vec\eta , \;\;\; \vec p = \epsilon^{1/2} t^{1/6} \vec\nu ,
\ee
the Wigner distribution takes the self-similar form
\be
f_{\rm W} = \epsilon^{-3/2} t^{-1/2} \int \frac{d\vec\eta\,'}{\pi^3} \, e^{2i\vec\eta\,' \cdot \vec\nu}
\hat\psi(\vec\eta-\vec\eta') \hat\psi^*(\vec\eta+\vec\eta') .
\ee
To take a self-similar form, the Husimi distribution must be derived from the Wigner distribution
by a smoothing that also follows the self-similar scaling $t^{-1/6}$ of spatial coordinates,
as in (\ref{eq:eta-def-x}). Choosing
\be
\sigma_x = 2^{-1/2} t^{-1/6} \sigma, \;\;\; \sigma_p = 1/(2\sigma_x) ,
\label{eq:sigma-x-sigma-p}
\ee
where as discussed for Eq.(\ref{eq:fH-positive}) we take 
$\sigma_x\sigma_p=1/2$ to obtain the Husimi distribution with the best possible
resolution while remaining positive, the self-similar Husimi distribution reads
\be
f_{\rm H}(\vec x, \vec p) = \epsilon^{-3/2} t^{-1/2} \hat f_{\rm H}(\vec\eta, \vec\nu) ,
\ee
with
\be
\hat f_{\rm H} = \left| \int \frac{d\vec \eta\,'}{2^{3/2} \pi^{9/4} \sigma^{3/2}}  \,
e^{-(\vec \eta - \vec \eta\,')^2/(2\sigma^2) - i \vec\nu \cdot \vec\eta\,'} \hat\psi(\vec\eta\,') \right|^2 .
\label{eq:fH-self-similar}
\ee
The parameter $\sigma$ sets the spatial resolution of this Husimi distribution.
At the background level, this gives
\be
\bar {\hat f}_{\rm H}(\vec\eta,\vec\nu) = \sigma^3 \pi^{-3/2} \, e^{-\sigma^2 \nu^2} .
\label{eq:fH-background}
\ee
The time dependence of the smoothing $\sigma_x$ gives rise to additional corrections
to the equation of motion followed by the Husimi distribution, as compared with the classical
Vlasov equation. However, this does not really matter, if we consider that the fundamental
distribution is the Wigner distribution, which satisfies equation (\ref{eq:Vlasov-Wigner})
that only deviates from the Vlasov equation by terms of order $\epsilon$ and higher.
Then, we are free to choose any smoothing for the Husimi distribution. This simply sets
the resolution that we choose and it can as well depend on time, following the growth
or the shrinking of the underlying dynamics.
Besides, these corrections are again of higher order over $\epsilon$,
so that for $\epsilon \to 0$ we can expect to recover the  classical dynamics on scales that are always
much larger than $\sigma_x$, unless small scales keep having a non-vanishing impact
on large scales.

\subsection{Linear regime}
\label{sec:linear-regime}

As for CDM, it is useful to study first small linear perturbations around the background,
which allows us to derive explicit analytical expressions. It also provides an interesting
comparison with the CDM case.
Linearizing the equations of motion (\ref{eq:continuity-2})-(\ref{eq:poisson-3})
in the density and velocity fields and combining as usual the continuity and Euler equations,
we obtain a closed equation for the linear density contrast $\delta_L$,
\be
\ddot{\delta}_L + \frac{4}{3t} \dot{\delta}_L - \frac{2}{3t^2}\delta_L + \frac{\epsilon^2}{4t^{8/3}}{\nabla^4_x}\delta_L =0 ,
\label{eq:linear-dynamics-1}
\ee
where the dot denotes the derivative with respect to the cosmic time $t$.
We recover the standard second-order equation in time also obtained for CDM, except for the
new last term associated with the quantum pressure, which comes with an $\epsilon$ prefactor as expected.
This term is negligible on large scales but damps short-scale modes because of its
Laplacian squared.

\subsubsection{Fourier space}

For comparison with the usual CDM scenario, it is useful to study first the equation
(\ref{eq:linear-dynamics-1}) in Fourier space, as this is the approach usually followed for CDM.
Then, Eq.(\ref{eq:linear-dynamics-1}) reads
\be
\ddot{\delta}_L + \frac{4}{3t} \dot{\delta}_L - \frac{2}{3t^2} \delta_L + \frac{\epsilon^2k^4}{4t^{8/3}}
\delta_L =0 .
\label{eq:deltaL-k}
\ee
As for CDM, different wavenumbers are decoupled and this second-order differential
equation has two independent solutions, $D_{\pm}(k,t)$, associated with the growing and decaying
modes
\ba
&& D_+(k,t) = t^{-1/6} J_{-5/2}\left(\frac{3}{2}\epsilon k^2 t^{-1/3}\right) , \nonumber \\
&& D_-(k,t) = t^{-1/6} J_{5/2}\left(\frac{3}{2}\epsilon k^2 t^{-1/3}\right) .
\label{eq:D_+-}
\ea
In the semiclassical limit, $\epsilon \to 0$, or on large scales, $k \to 0$, we recover the
time dependence of the CDM linear growing modes, $D_+(k,t) \propto t^{2/3} \propto a$
and $D_-(k,t) \propto t^{-1}$.
However, in contrast with the CDM modes, for nonzero $\epsilon$ the modes $D_{\pm}(k,t)$
depend on the wavenumber $k$ and differ from power laws.
At high $k$ or at small time, we have
$D_+(k,t) \sim \cos(3 \epsilon k^2 t^{-1/3}/2)$ and
$D_-(k,t) \sim \sin(3 \epsilon k^2 t^{-1/3}/2)$. Thus, we obtain acoustic waves when the quantum
pressure is dominant (but with a higher power of $k$ because of the $k^4$ factor
in (\ref{eq:deltaL-k})).
At late times we always recover the CDM behavior. This is due to the damping of the
quantum-pressure term in Eq.(\ref{eq:deltaL-k}) by the factor $t^{-8/3}$.
This is also related to the scalings $r \propto \sqrt{t}$ and $x \propto t^{-1/6}$ found in
(\ref{eq:self-r-x}): the scale where wavelike effects, or the quantum pressure, are important
decreases in time, in comoving coordinates. Thus, deviations from CDM become confined
to increasingly small scales.

Then, the linear density contrast takes the form
\be
\delta_L(\vec k,t) = D_+(k,t) \delta_{L+}(\vec k) + D_-(k,t) \delta_{L-}(\vec k) ,
\label{eq:delta-L-D+-}
\ee
where the functions $\delta_{L\pm}(\vec k)$ are determined by the initial conditions
for $\delta$ and $\dot\delta$ at some initial time.
From Eq.(\ref{eq:self-r-x}) we find that spherically symmetric and self-similar solutions
are of the form
\be
\delta(\vec k,t) = t^{-1/2} \, \delta(t^{-1/6} k) .
\ee
Comparing with Eqs.(\ref{eq:D_+-}) and (\ref{eq:delta-L-D+-}), we obtain
$\delta_{L\pm}(\vec k) \propto k^{-2}$.
We require $\delta_L(k) \to 0$ for $k \to 0$, as we wish to recover the background density
on large scales. This rules out the mode $D_+(k,t)$ and we obtain
\be
\delta_L(\vec k,t) \propto t^{-1/6} k^{-2}  J_{5/2}\left(\frac{3}{2}\epsilon k^2 t^{-1/3}\right) .
\label{eq:deltaL-decaying}
\ee
Note that in the FDM regime, where the quantum pressure is important, the two modes $D_\pm$
oscillate with a constant amplitude and no longer correspond to growing and decaying modes.
Therefore, it is not unphysical to keep only the mode $D_-$, as a small perturbation by 
a mode $D_+$ will remain small as long as we remain in the FDM regime.
Going back to real space by taking the inverse Fourier transform,
we obtain
\be
\delta_L(x,t) = 1+ \frac{\eta^4}{45} - \frac{8\eta^2}{9\pi}  \, _2F_3\left(-\frac{1}{2},2;\frac{3}{2},\frac{5}{4},\frac{7}{4};-\frac{\eta ^4}{144}\right) ,
\label{eq:delta-L-1-2F3}
\ee
where $\,_2F_3$ is a hypergeometric function, $\eta$ is the scaling variable defined in
(\ref{eq:eta-def-x}), and we normalized the linear mode to unity at the center.

As already explained below (\ref{eq:self-r-x}), this self-similar solution expands in physical
coordinates $\vec r$ but shrinks in comoving coordinates $\vec x$.
Another difference with the CDM self-similar solutions is that the amplitude of the linear
density contrast near the center does not grow with time and remains constant.
Thus, it is not unstable and does not reach the nonlinear regime at late times:
a small-amplitude perturbation $\delta_L$ will always keep a small amplitude
and to reach the nonlinear regime we must start with a large nonlinear perturbation.

\subsubsection{Real space}
\label{sec:real-space-linear}

Because we look for spherically-symmetric self-similar solutions of the form (\ref{eq:self-r-x}),
we can actually solve the equation of motion (\ref{eq:linear-dynamics-1}) in real space.
Looking for a solution $\delta(\eta)$, in terms of the self-similar scaling variable (\ref{eq:eta-def-x}),
the partial differential equation (\ref{eq:linear-dynamics-1}) becomes the ordinary differential
equation
\be
\delta_L^{(4)} + \frac{4}{\eta}\delta_L^{(3)} + \frac{\eta^2}{9}\delta_L'' + \frac{\eta}{3}\delta_L'
- \frac{8}{3}\delta_L  = 0 ,
\label{eq:linear-deltaL}
\ee
where the prime denotes the derivative with respect to $\eta$.

The fourth and third derivatives come from the quantum pressure term, which changes
the order of the equation from two to four, as compared with the usual CDM case.
Therefore, we now have four independent linear modes instead of two, which read
\ba
&& \delta_{L1} = 45 + \eta^4 , \;\; \delta_{L2} = \frac{1}{\eta}  \,
_2F_3\left(-\frac{5}{4},\frac{5}{4};\frac{1}{4},\frac{1}{2},\frac{3}{4};-\frac{\eta ^4}{144}\right) ,
\nonumber \\
&& \delta_{L3} = \eta  \;
_2F_3\left(-\frac{3}{4},\frac{7}{4};\frac{3}{4},\frac{5}{4},\frac{3}{2};-\frac{\eta ^4}{144}\right) ,
\nonumber \\
&& \delta_{L4} = \eta^2  \;
_2F_3\left(-\frac{1}{2},2;\frac{5}{4},\frac{3}{2},\frac{7}{4};-\frac{\eta ^4}{144}\right) .
\label{eq:linear-1-4}
\ea
Their asymptotic behaviors at the center read
\be
\eta\to 0 : \;\;  \delta_{L2} = \frac{1}{\eta} + \dots , \;\; \delta_{L3} = \eta + \dots ,
\;\; \delta_{L4} = \eta^2 + \dots
\ee
where the dots stand for higher-order terms.
This rules out $\delta_{L2}$ and $\delta_{L3}$ if we look for a smooth solution with an even
Taylor expansion in the radius $x$ at the center.
The asymptotic behavior of $\delta_{L4}$ at large distance reads
\ba
&& \hspace{-0.3cm}  \eta\to \infty : \;\; \delta_{L4} = \frac{\pi \eta^4}{40} + \frac{9\pi}{8} - \frac{243}{5\eta^6} + \dots
+ \cos(\eta^2/6)  \nonumber \\
&&  \times \left[ \frac{27 \sqrt{3\pi}}{2\eta^3} + \dots \right]
+ \sin(\eta^2/6) \left[ \frac{27 \sqrt{3\pi}}{2\eta^3} + \dots \right]
\label{eq:delta-L4-large-eta}
\ea
where the dots stand for higher order terms in $1/\eta$.
Therefore, the only combination of the four modes that satisfies the boundary conditions
at the center and at infinity is
\be
\delta_L = - \frac{8}{9\pi} \left( \delta_{L4} - \frac{\pi}{40} \delta_{L1} \right) ,
\label{eq:deltaL-deltaL4-deltaL1}
\ee
where we chose the normalization $\delta_L(0)=1$,
and we recover Eq.(\ref{eq:delta-L-1-2F3}), as expected.
From the density contrast $\delta_L$ we can derive the velocity $u_L$ and the perturbed mass
$\delta M_L$, which can also be expressed in terms of hypergeometric functions.

\begin{figure}[h]
\centering
\includegraphics[width=0.49\textwidth]{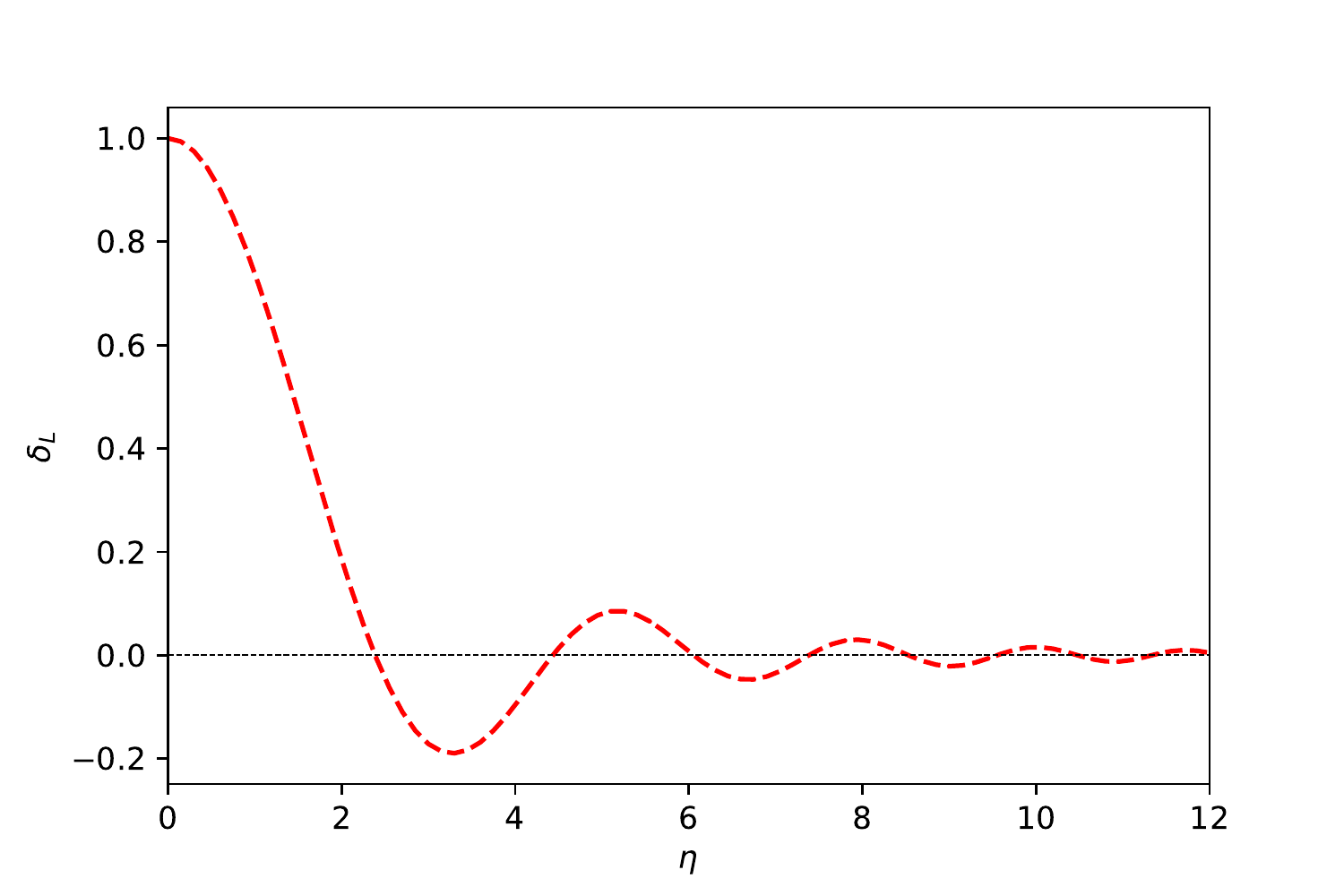}
\includegraphics[width=0.49\textwidth]{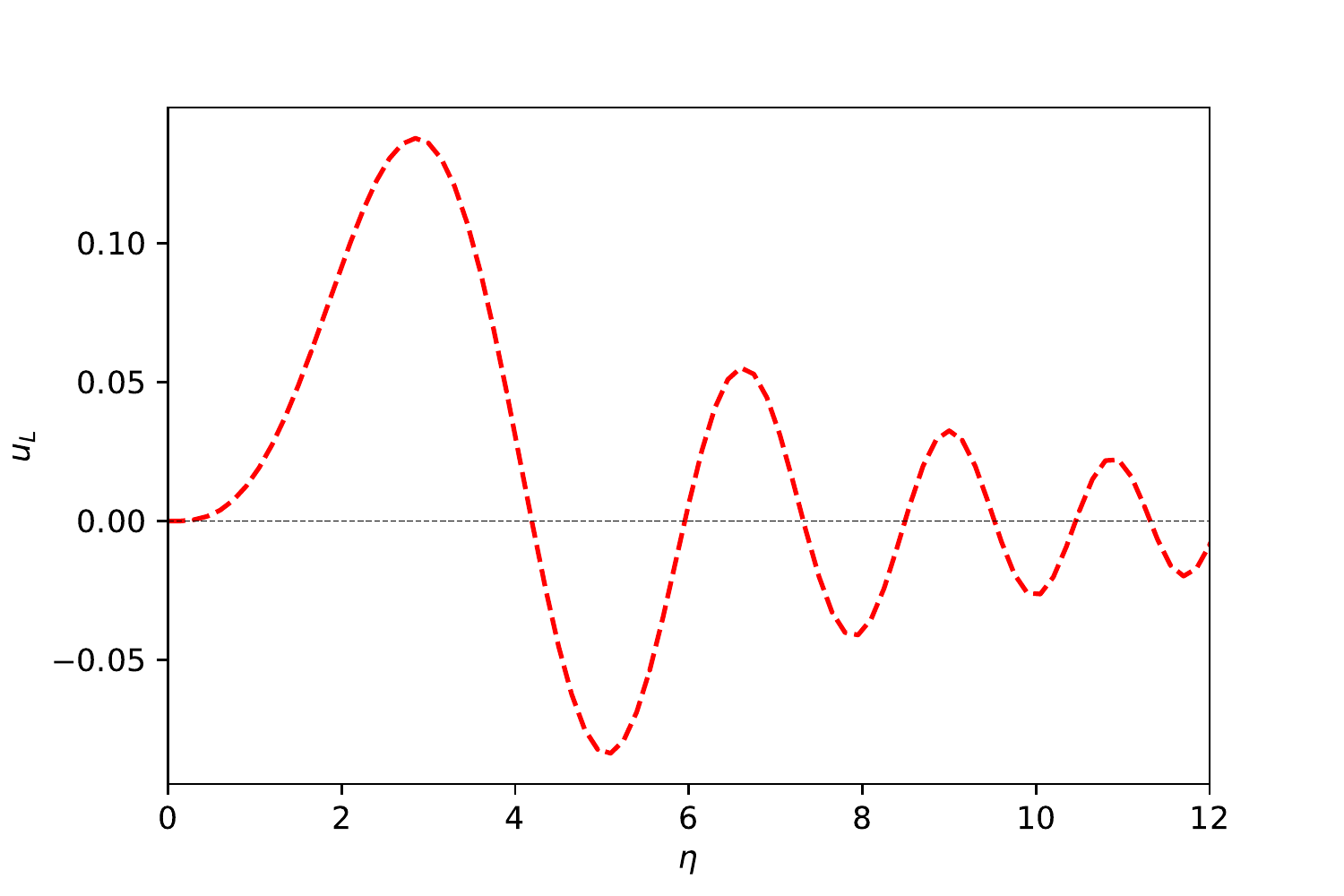}
\includegraphics[width=0.49\textwidth]{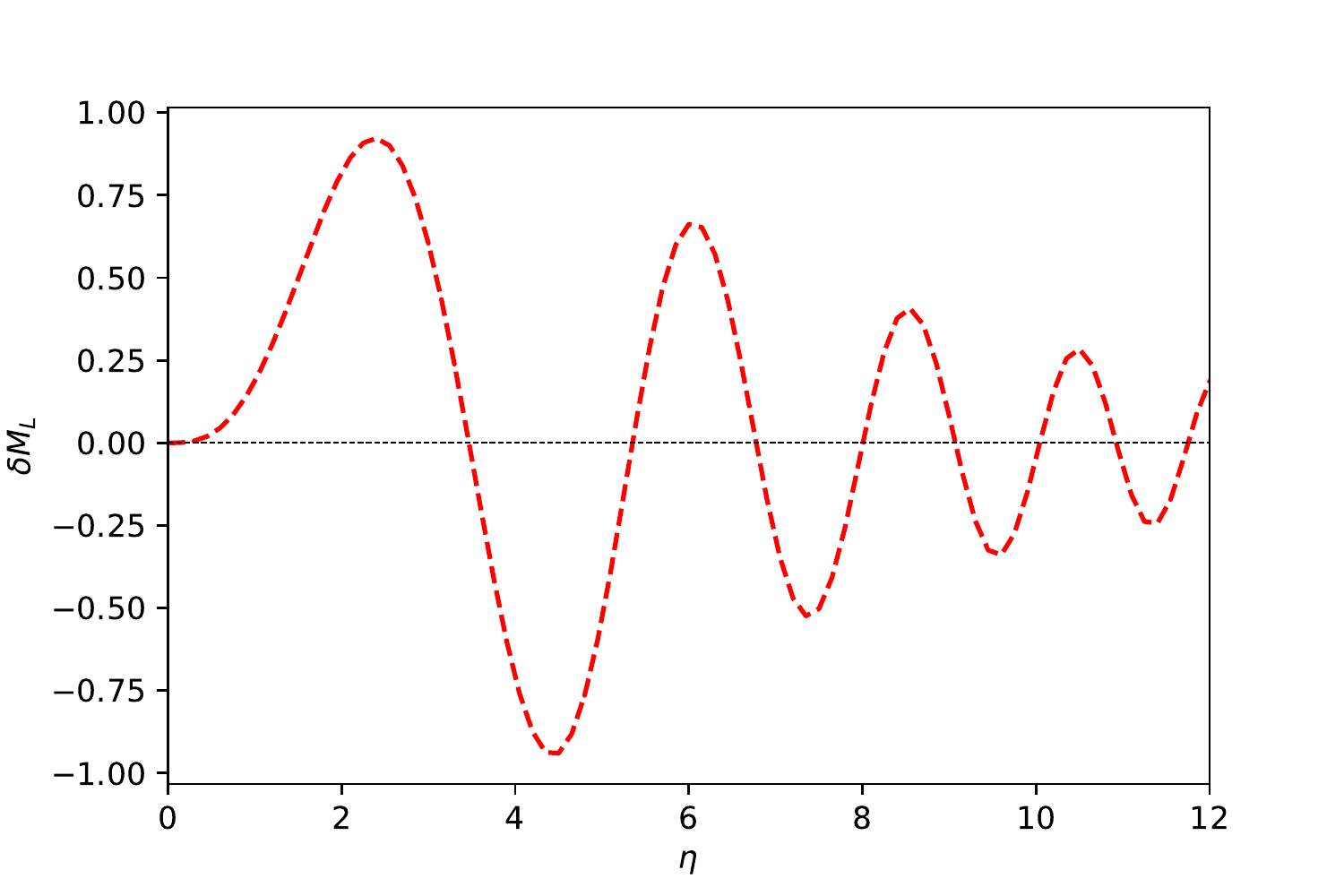}
\caption{From top to bottom: linear density contrast $\hat\delta_L$ from Eq.(\ref{eq:delta-L-1-2F3}),
linear velocity perturbation $\hat u_L$ and linear mass perturbation $\delta\hat M_L$,
for $\hat\delta_L(0)=1$.}
\label{fig:linear-result}
\end{figure}

\subsubsection{Numerical results}
\label{sec:numerical-linear}

We show in Fig.~\ref{fig:linear-result} the linear density contrast, velocity and mass perturbations,
normalized to $\hat\delta_L(0)=1$.
In contrast with CDM self-similar solutions, which show a power-law falloff at large distance
without oscillations, the fields oscillate around zero at the same frequency.
In particular, we can see that these self-similar solutions correspond to compensated profiles,
as the mass perturbation goes to zero at infinity and actually changes sign an infinite number
of times.
Positive density peaks approximately correspond to minima of the radial velocity perturbation and to
zero-crossings of the mass perturbation.
Of course, these oscillatory features are due to the quantum pressure (absent for CDM)
which generates acoustic waves (but with a different dispersion relation than usual sound
waves).
We can see that the central density peak is much higher than the next peaks, which decrease
as $1/\eta^3$ at large distance. This is due to the dimension three of space (i.e. the volume
factor $\eta^3$) as the velocity and mass oscillations appear more regular.

As time grows, the profiles shrink in comoving coordinates as $x \propto t^{-1/6}$
from Eq.(\ref{eq:eta-def-x}). The central peak becomes confined to an increasingly narrow
region in comoving coordinates, with a decreasing mass. However, it is actually expanding
in physical coordinates as $r \propto \sqrt{t}$.

In contrast with CDM solutions, the amplitude of the linear density-contrast profile does
not evolve with time. However, the velocity and mass perturbations decay as $1/\sqrt{t}$,
see the scaling laws (\ref{eq:self-r-x}).

\subsubsection{Balance of kinetic, gravitational and quantum-pressure terms}
\label{sec:balance-linear}

While the static solitons (\ref{eq:soliton-eq}) are governed by the balance between gravity and
the quantum pressure in (\ref{eq:hydrostatic}), the self-similar solutions are dynamical and involve
in addition kinetic effects, associated with the nonzero velocity, see (\ref{eq:Bernoulli}).
In fact, at large distance we obtain for $\eta \to \infty$
\ba
&& \hat u_L \sim \eta^{-5} + \eta^{-2} [ \cos(\eta^2)/6+ \sin(\eta^2)/6) ]  , \nonumber \\
&& \hat\varphi_{{\rm N} L} \sim \eta^{-4} + \eta^{-5} [ \cos(\eta^2)/6+ \sin(\eta^2)/6) ]  ,  \nonumber \\
&& \hat\Phi_{{\rm Q}L} \sim \eta^{-8} + \eta^{-1}  [ \cos(\eta^2)/6+ \sin(\eta^2)/6) ] ,
\label{eq:large-eta-linear}
\ea
where we only wrote the leading smooth and oscillatory terms, omitting the numerical
factors. Thus, we find that at large distance the Bernoulli equation (\ref{eq:Bernoulli}) is
governed by the balance between the kinetic energy and the quantum pressure, while
gravity becomes negligible.
This is possible thanks to the oscillations of the perturbed mass $\delta\hat M$ around zero,
somewhat like a compensated density profile. This entails a screening of the central
density peak and of its gravitational attraction, which makes gravity negligible at large radius.
This is not totally surprising. Gravity is a long-range force, as shown by the inverse Laplacian
in the Poisson equation (\ref {eq:poisson-2}), whereas the quantum pressure is a short-range
force, as shown by the Laplacian in Eq.(\ref{eq:quantum-pressure}).
This implies that the quantum pressure cannot balance gravity at large distance.
Therefore, either gravity is balanced by the kinetic energy (as for the Hubble flow in the
Einstein-de Sitter background), or it is screened by the compensated density profile.
In the latter case, the small residual gravity can be balanced by the quantum pressure,
but in our solution (\ref{eq:large-eta-linear}) the screening is so efficient that at large
distance we simply have free waves, with a balance between the quantum pressure
and kinetic terms, on top of the cosmological background.
Of course, at the center where the velocity vanishes by symmetry we have instead a balance
between gravity and the quantum pressure.
Therefore, we have a change of the nature of the dynamics with the radius.
It is set by gravity vs quantum pressure at the center and by kinetic terms vs quantum pressure
at large distance.


\subsection{Comparison with CDM}
\label{sec:CDM-linear}

Already at the linear level, we can see that the FDM self-similar solutions
are very different from the CDM ones. 
First, the CDM linear modes $D_\pm(t)$ are scale-independent with $D_+(t) \propto t^{2/3}$
and $D_-(t) \propto t^{-1}$. 
We recover these power laws in the small-$\epsilon$ limit of Eq.(\ref{eq:D_+-}).
Since for CDM the space and time dependences factorize in Eq.(\ref{eq:delta-L-D+-}),
requirements on the shape of the density profile at an initial time do not rule out the
growing nor the decaying mode. Then, one usually only keeps the growing mode,
assuming that the decaying mode has had time to become negligible.
In contrast, in Eq.(\ref{eq:deltaL-decaying}) we only kept the mode $D_-$
because of the requirement to converge to the cosmological background on large scales.
However, as we noticed below Eq.(\ref{eq:deltaL-decaying}), for FDM the linear modes 
$D_\pm$ are no longer growing and decaying modes but acoustic oscillations of similar amplitude.
Therefore, the self-similar solution associated with the mode $D_-$ is physical,
as a small perturbation associated with the mode $D_+$ will remain small as long as we remain
in the FDM regime where the quantum pressure is important.
As seen in  Eq.(\ref{eq:large-eta-linear}), this is valid at large radii in the linear regime.
In fact, as seen in the lower panels in Figs.~\ref {fig:non-linear-10} and \ref{fig:non-linear-100}
below, this is valid at all radii at the nonlinear level for the self-similar solutions studied in this paper.

As the Newtonian equations of motion (\ref{eq:continuity-2})-(\ref{eq:poisson-3})
only apply to sub-Hubble scales, one may consider introducing a large-scale
cutoff at the Hubble radius, so that the FDM mode $D_+(k,t)$ does not lead
to divergent quantities. Then, one may wonder whether one could recover
the CDM self-similar solutions by first taking the semiclassical limit
$\epsilon\to 0$ and next $c/H_0 \to \infty$.
This is better discussed in real space, as we detail below.

First, let us recall the main properties of the CDM self-similar solutions for the 
Einstein-de Sitter cosmology \citep{Fillmore:1984wk}, associated with overdensities
and the formation of spherical virialized halos.
For an initial overdense power-law profile
\be
0 < \gamma < 3 : \;\;\; \delta_L(r) \propto r^{-\gamma} 
\label{eq:CDM-def}
\ee
of the linear density contrast, one obtains nonlinear self-similar solutions, 
with a turnaround radius that grows with time as
\be
r_{\rm ta}(t) \propto t^{2/3+2/(3\gamma)} 
\label{eq:CDM-r-ta}
\ee
and a nonlinear density profile in the inner virialized regions, for $r \ll r_{\rm ta}$,
that shows a different power law,
\be
\gamma \leq 2: \;\; \rho \propto r^{-2} , \;\;\; \gamma \geq 2 : \;\ 
\rho \propto r^{-3\gamma/(1+\gamma)} .
\label{eq:CDM-nonlinear}
\ee
For shallow initial profiles, $\gamma < 2$, the mass within a small radius $R$ is dominated 
by the particles that have just collapsed, whereas for steep initial profiles, $\gamma > 2$, 
the mass within $R$ is dominated by the particles that have collapsed long ago, when the 
turnaround radius was of the order of $R$  \citep{Fillmore:1984wk}.
These solutions, which exhibit gravitational instability and collapse of increasingly massive
and distant shells, originate from the growing mode $D_+(t)$ in the linear regime,
with the power-law radial profile (\ref{eq:CDM-def}).

For FDM, the mode $D_+$ in (\ref{eq:delta-L-D+-}) would correspond to a $k^{-7}$
profile in the semiclassical limit, which we discarded because of the low-$k$ divergence.
By power counting, this corresponds to an $x^4$ profile in real space.
This is indeed recovered in the real-space analysis of Sec.~\ref{sec:real-space-linear},
in the linear mode $\delta_{L1}$ of Eq.(\ref{eq:linear-1-4}), which grows as $\eta^4$ at large
$\eta$ [from Eq.(\ref{eq:eta-def-x}) we can check that the semiclassical limit
$\epsilon\to 0$ corresponds to $\eta\to\infty$].
Comparing with Eq.(\ref{eq:CDM-def}), this corresponds at the
formal level to $\gamma=-4$ and Eq.(\ref{eq:CDM-r-ta}) gives for the characteristic scale
at time $t$ the power law $r_{\rm ta} \propto \sqrt{t}$.
Thus, we recover the square-root growth in physical coordinates obtained in
Eq.(\ref{eq:eta-def-x}).

Thus, for CDM the self-similar exponents, such as the growth of the characteristic scale
in Eq.(\ref{eq:CDM-r-ta}), span a continuous range, indexed on the slope $\gamma$ of the
initial profile of the density contrast.
In FDM, in addition to gravity we have the new quantum pressure in the hydrodynamical
equations of motion. As can be expected, this new term restricts the possibility of self-similar
solutions. Because it takes a power-law form, it still permits the existence of self-similar
solutions but with a unique exponent, associated with the square-root growth of the physical
scale in Eq.(\ref{eq:eta-def-x}). As such, it selects among the CDM solutions
(\ref{eq:CDM-def})-(\ref{eq:CDM-nonlinear}) the one obtained for $\gamma=-4$,
as this is the only exponent compatible with the quantum pressure term.
However, this only holds at a formal level, because this value of the exponent $\gamma$
is beyond the allowed range in Eq.(\ref{eq:CDM-def}).
In this sense, the semiclassical limit would make this FDM self-similar solution converge to
a specific CDM self-similar solution (the one with the same exponent $\gamma$),
but this is not possible or relevant in practice, because this specific solution is actually
badly behaved (for both CDM and FDM) because it does not converge to the background
on large scales.

Therefore, the allowed FDM self-similar solution, which we study in more details in this paper,
does not correspond to a standard CDM self-similar solution.
In fact, it would correspond to a different CDM self-similar solution associated with a decaying mode.
For CDM, such a solution is not of great practical interest, as one expects to be dominated
by growing modes. 
In contrast, as we noticed above, for FDM the linear modes 
$D_\pm$ are no longer growing and decaying modes but acoustic oscillations of similar amplitude.
Therefore, the self-similar solution associated with the mode $D_-$ is physical,
as a small perturbation associated with the mode $D_+$ remains small.
In fact, we shall see in Sec.~\ref{sec:nonlinear-regime} that nonlinearities remain important
at all scales, in the sense that the self-similar solution does not converge to the linear-theory
prediction (\ref{eq:deltaL-deltaL4-deltaL1}) at large radii, because it includes additional contributions
associated with the modes $\delta_{L2}$ and $\delta_{L3}$.

Even though it does not have a standard CDM counterpart, it is still interesting to study this 
FDM self-similar solution for its own sake. 
It permits an analytical or semi-analytical treatment beyond static solitons and it provides 
an explicit example of gravitational cooling.
Thus, it can help understand the dynamics generated by the competition between 
the quantum pressure, gravity and kinetic effects.

\subsection{Nonlinear regime}
\label{sec:nonlinear-regime}

\subsubsection{Closed equation over $\delta M$}

We now turn to the nonlinear regime and look for exact self-similar solutions of the equations
of motion (\ref{eq:continuity-2})-(\ref{eq:poisson-3}) of the form (\ref{eq:self-r-x}).
In terms of the self-similar coordinate $\eta$, the Poisson equation reads
\be
\frac{1}{\eta^2} \frac{d}{d\eta} \left( \eta^2 \frac{d\hat\varphi_{\rm N}}{d\eta} \right) = \frac{2}{3} \hat\delta ,
\ee
while the quantum pressure reads
\be
\hat\Phi_{\rm Q} = - \frac{1}{2 \eta^2 \sqrt{1+\hat\delta}} \frac{d}{d\eta} \left( \eta^2 \frac{d}{d\eta}
\sqrt{1+\hat\delta} \right) .
\label{eq:Phi-Q-eta}
\ee
The density contrast is given by the first derivative of the mass perturbation,
\be
\hat\delta = \frac{3}{2\eta^2} \delta\hat M ' .
\label{eq:delta-deltaM}
\ee
Integrating the continuity equation once over the radial coordinate gives the expression of the
radial velocity in terms of the perturbed mass as
\be
\hat u = \frac{3 \delta\hat M - \eta \delta\hat M'}{4\eta^2+6\delta\hat M'} .
\ee
Substituting these expressions into the Euler equation (\ref{eq:Euler-eta}),
we obtain a closed nonlinear equation for the perturbed mass $\delta\hat M$,
\ba
&& \hspace{-0.4cm}
9 ( 2 \eta^3 + 3 \eta \delta\hat M' )^2 \delta\hat M^{(4)} - ( 144 \eta^5 + 216 \eta^3 \delta\hat M'
+ 108 \eta^4 \delta\hat M''      \nonumber \\
&&  \hspace{-0.2cm}
+ 162 \eta^2 \delta\hat M' \delta\hat M'' ) \delta\hat M^{(3)} + ( 4 \eta^8 + 288 \eta^4
+ 36 \eta^5 \delta\hat M      \nonumber \\
&&  \hspace{-0.2cm}
- 216 \eta^2 \delta\hat M' + 324 \eta^3 \delta\hat M''+ 81 \eta^2 \delta\hat M^2
+ 81 \eta^2 \delta\hat M''^2 ) \delta\hat M''      \nonumber \\
&&  \hspace{-0.2cm}
- 3 ( 4 \eta^7 + 96 \eta^3 + 180 \eta^4 \delta\hat M + 243 \eta^2 \delta\hat M \delta\hat M'
- 3 \eta^3 \delta\hat M'^2      \nonumber \\
&&  \hspace{-0.2cm}
+ 108 \delta\hat M \delta \hat M'^2 ) \delta \hat M' - 12 \eta^3 (7 \eta^3 - 9 \delta \hat M )
\delta\hat M = 0 .
\label{eq:deltaM-nonlinear}
\ea

\subsubsection{Comparison with the linear equation}

If we linearize Eq.(\ref{eq:deltaM-nonlinear}) we obtain the fourth-order linear equation
\ba
\mbox{(L1)} : && 9 \eta^3 \delta\hat M^{(4)} - 36 \eta^2 \delta\hat M^{(3)}
+ (72 \eta + \eta^5) \delta\hat M''       \nonumber \\
&& -3 (24 + \eta^4) \delta\hat M'  - 21 \eta^3 \delta\hat M =  0 ,
\label{eq:L1}
\ea
whereas from the linear equation (\ref{eq:linear-deltaL}), using (\ref{eq:delta-deltaM}) we
obtain the fifth-order linear equation
\ba
\mbox{(L2)} : && 9 \eta^4 \delta\hat M_L^{(5)} - 36 \eta^3 \delta\hat M_L^{(4)} + (108 \eta^2
+ \eta^6) \delta\hat M_L^{(3)}  \nonumber \\
&&  - (216 \eta + \eta^5) \delta\hat M_L'' + 24 (9 - \eta^4) \delta\hat M_L' =  0 .
\ea
As it should, we can check that these two equations are related,
\be
\mbox{(L2)} = \eta^4 \frac{d}{d\eta} \left[ \eta^{-3} \mbox{(L1)} \right] .
\ee
Thus, the solutions of (L1) are also solutions of (L2). The latter being of order five instead of
four, it includes the additional solution $\mbox{(L1)} \propto \eta^3$, corresponding to
$\delta M_L = {\rm constant}$, and hence to $\delta_L=0$.
Therefore, the linearized equation (\ref{eq:L1}) is fully consistent with the linear theory
studied in Sec.~\ref{sec:linear-regime}.

\subsubsection{Numerical procedure}

We solve the nonlinear equation (\ref{eq:deltaM-nonlinear}) with a shooting method \citep{Press1992},
subdividing the spatial domain $\eta \geq 0$ in three regions: a central region
$\eta \lesssim 0.1$, an intermediate region $0.1 \lesssim \eta \lesssim 10$, and a large-distance
region $\eta \gtrsim 10$.
This allows for a convenient implementation of the boundary conditions.

We look for solutions that converge to the cosmological background at large distance,
so that the density contrast and the perturbed mass go to zero.
Therefore, at large distances we can use the linearized equation (\ref{eq:L1}).
This gives the four independent linear modes,
\be
\begin{split}
& \delta\hat M_{L1} = \eta ^3 \left( 105 + \eta ^4 \right) , \\
& \delta\hat M_{L2} = \, _2F_3 \left( -\frac{7}{4}, \frac{3}{4}; -\frac{1}{4}, \frac{1}{4}, \frac{1}{2};
-\frac{\eta ^4}{144} \right)  + \frac{\sqrt{\pi}}{180\sqrt{3}} \delta\hat M_{L1} , \\
& \delta\hat M_{L3} = \eta ^2 \, _2F_3 \left( -\frac{5}{4}, \frac{5}{4}; \frac{1}{4}, \frac{3}{4},
\frac{3}{2}; -\frac{\eta ^4}{144}\right) - \frac{\sqrt{\pi}}{126\sqrt{3}} \delta\hat M_{L1}  , \\
& \delta\hat M_{L4} = \eta ^5 \, _2F_3 \left( -\frac{1}{2}, 2; \frac{3}{2}, \frac{7}{4}, \frac{9}{4};
-\frac{\eta ^4}{144}\right) - \frac{\pi }{56} \delta\hat M_{L1} .
\end{split}
\ee
Therefore, at large distance the perturbed mass is a combination of these four modes,
\be
\eta \to \infty : \;\;\delta\hat M = \sum_{i=1}^4 c_i \, \delta\hat M_{Li}  ,
\label{eq:deltaM-linear-c_i}
\ee
whith coefficients $c_i$ to be determined.
For $\eta\to\infty$ the linear modes show the large-distance behaviors
\ba
&& \hspace{-0.3cm} \delta\hat M_{L2} = \cos(\eta^2/6) \left[ \frac{2079}{\eta^4} + \dots \right]
- \sin(\eta^2/6) \left[ \frac{63}{\eta^2} + \dots \right] , \nonumber \\
&& \hspace{-0.3cm} \delta\hat M_{L3} = - \cos(\eta^2/6) \left[ \frac{90}{\eta^2} + \dots \right]
- \sin(\eta^2/6) \left[ \frac{2970}{\eta^4} + \dots \right] , \nonumber \\
&& \hspace{-0.3cm} \delta\hat M_{L4} = - \cos(\eta^2/6) \left[ \frac{405 \sqrt{3\pi}}{2\eta^2}
+ \dots \right]    \nonumber \\
&& \hspace{0.8cm} + \sin(\eta^2/6) \left[ \frac{405 \sqrt{3\pi}}{2\eta^2} + \dots \right] .
\ea
Therefore, we can see that the three modes $\delta\hat M_{L2}$, $\delta\hat M_{L3}$ and
$\delta\hat M_{L4}$ obey the boundary condition $\delta\hat M \to 0$,
while the divergent contribution from the mode $\delta\hat M_{L1}$ is ruled out.
This gives the large-distance boundary condition
\be
c_1 = 0 .
\label{eq:c_1=0}
\ee

To implement the boundary condition at the center, we write the Taylor expansion
\be
\eta \to 0 : \;\;\; \delta\hat M = a_3 \eta^3 + a_5 \eta^5 + a_7 \eta^7 + \dots ,
\label{eq:deltaM-center}
\ee
corresponding to a density contrast with is finite and smooth at the center, with an expansion
in even powers of $x$.
In particular, the density contrast at the origin is
\be
\hat\delta(0) = 9 a_3/2 .
\ee
Thus, the central density contrast $\hat\delta(0)$
specifies the value of $a_3$. Next, substituting the expansion (\ref{eq:deltaM-center}) into the
differential equation (\ref{eq:deltaM-nonlinear}) gives a hierarchy of equations that determines
all higher-order coefficients $\{ a_7, a_9, ... \}$ in terms of $\{a_3,a_5\}$.
Thus, we are left with only one free parameter $a_5$, which is set by the boundary condition
(\ref{eq:c_1=0}) at infinity.

In practice, we first choose the value of $\hat\delta(0)$, hence of $a_3$, of the profile we aim
to compute. Then, for a trial value of $a_5$, we compute the profile $\delta\hat M(\eta)$
up to $\eta_- \sim 0.1$ with the Taylor expansion (\ref{eq:deltaM-center}), all higher-order
coefficients being known from the substitution into the differential equation
(\ref{eq:deltaM-nonlinear}). Next, we advance up to $\eta_+ \sim 10$ by solving the
nonlinear differential equation (\ref{eq:deltaM-nonlinear}) with a Runge-Kutta algorithm.
Then, at $\eta_+$, far enough in the linear regime, we match to the linear expansion
(\ref{eq:deltaM-linear-c_i}).
For a random initial guess $a_5$ at the center, this will give a nonzero coefficient $c_1$.
Therefore, we use an iterative scheme over $a_5$ until the matching coefficient $c_1$
at the outer boundary $\eta_+$ vanishes.
This sets the value of $a_5$.

In general, all three coefficients $c_2, c_3$ and $c_4$ in the linear expansion
(\ref{eq:deltaM-linear-c_i}) are nonzero.
In contrast, the linear-theory solution (\ref{eq:deltaL-deltaL4-deltaL1}) actually corresponds to
the linear mode $\delta\hat M_{L4}$ only, with $c_2=c_3=0$.
This corresponds to the fact that the linear modes $\delta\hat M_{L2}$ and $\delta\hat M_{L3}$
do not satisfy the appropriate boundary conditions at the center, $\eta\to 0$,
which only leaves $\delta\hat M_{L4}$ as the unique solution (up to a normalization) for a linear
solution valid over the full range $0 \leq \eta < \infty$.
This agrees with the fact that we obtained only one linear solution in Sec.~\ref{sec:linear-regime}.

Once we consider the exact nonlinear differential equation (\ref{eq:deltaM-nonlinear}), matters
are different and the modes $\delta\hat M_{L2}$ and $\delta\hat M_{L3}$ show nonzero
contributions at large distances.
This is because at the center the perturbations do not asymptotically vanish (contrary to what
happens at $\eta\to \infty$) but remain finite, so that nonlinear contributions cannot be fully
neglected. This changes the mapping between the boundary conditions at the center and
at infinity and the large-distance modes $\delta\hat M_{L2}$ and $\delta\hat M_{L3}$
are no longer excluded because the behavior at the center of the linear modes is no longer
relevant, as this central region is beyond strict linear theory.
However, for $\hat\delta(0) \to 0$ the contributions from the modes $\delta\hat M_{L2}$ and
$\delta\hat M_{L3}$ become small as compared with that from $\delta\hat M_{L4}$,
as the solution converges to the linear theory.

Therefore, in contrast with the CDM self-similar solutions, for a finite central density
there is  never convergence to the linear theory at large distance, in the sense that $\delta\hat M(\eta)$
does not converge to $\delta\hat M_L(\eta) \propto \delta\hat M_{L4}(\eta)$.
Indeed, the additional modes $\delta\hat M_{L2}$ and $\delta\hat M_{L3}$
also decrease as $1/\eta^2$, with oscillatory prefactors.
This is due to a strong coupling between the behaviors at the center and at infinity.
This arises from the self-similarity of the solution and from the quantum pressure
(absent for CDM) which propagates information from the center to infinity and vice versa
(by looking for a self-similar solution we have implicitly provided an infinite amount of time to acoustic waves
to propagate over all space).
Physically, the blow-up character of the solutions means that the scalar matter starts in the nonlinear regime at small radii, and ends in the linear regime (i.e. converges to the Hubble flow) at late times
and large radii, as explicitly seen in Sec.~\ref{sec:trajectories}.
This is the opposite of the usual CDM collapsing solutions, where matter shells start in the linear regime and finally collapse and virialize in the nonlinear inner regions.
This couples the final linear era to the initial nonlinear era.

\subsection{Overdensities}
\label{sec:overdensities}

\begin{figure}[h!]
\centering
\includegraphics[width=0.48\textwidth]{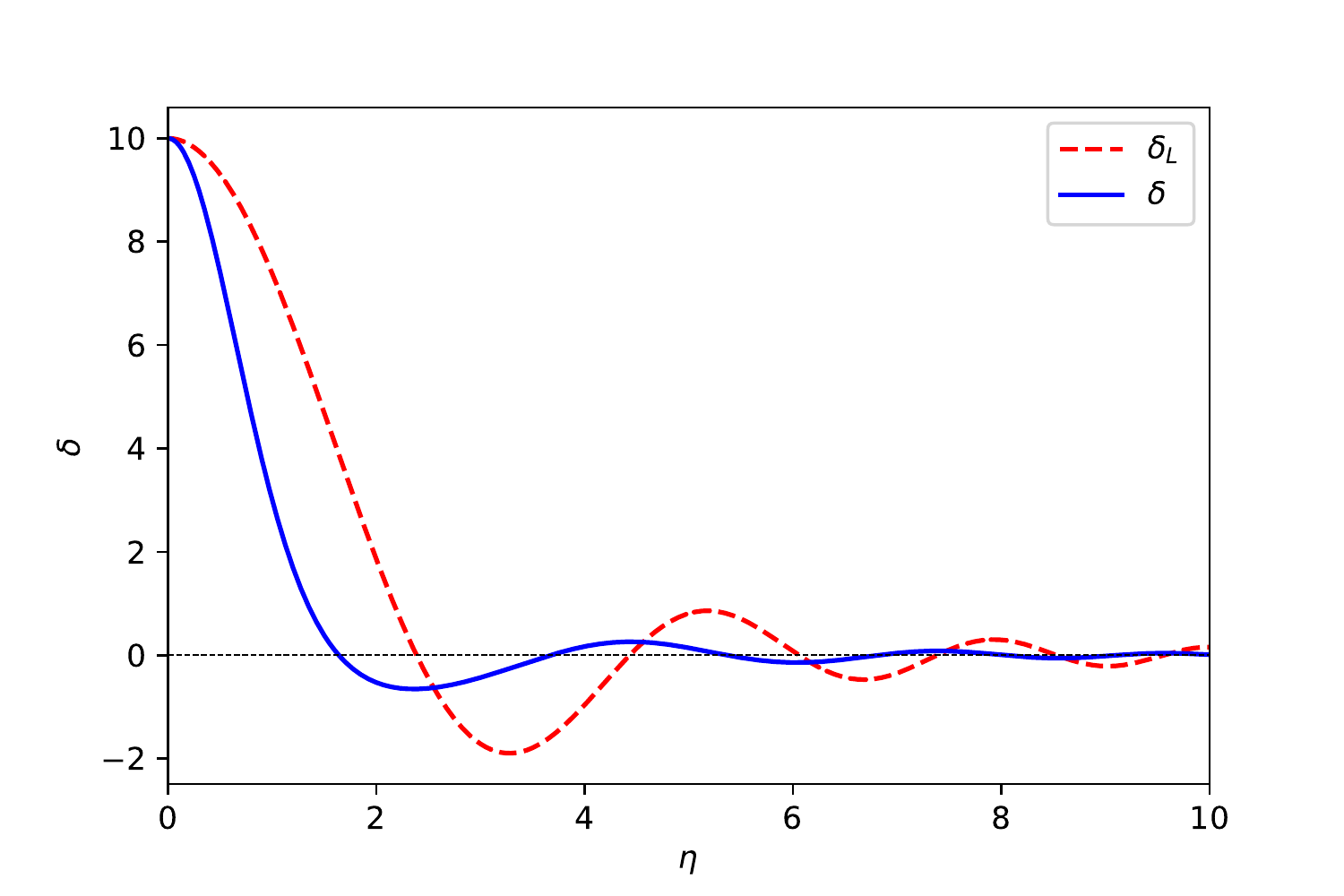}
\includegraphics[width=0.48\textwidth]{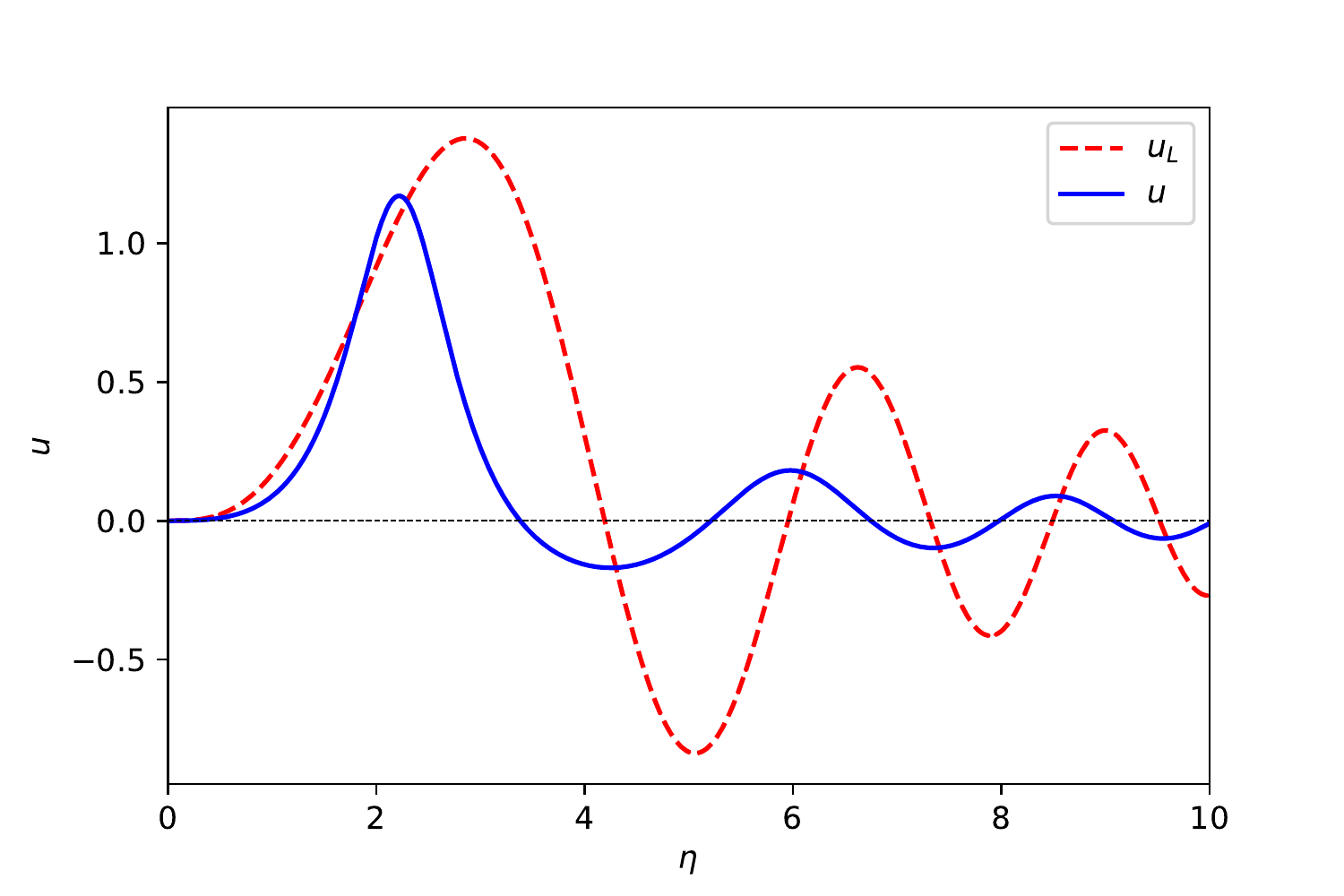}
\includegraphics[width=0.49\textwidth]{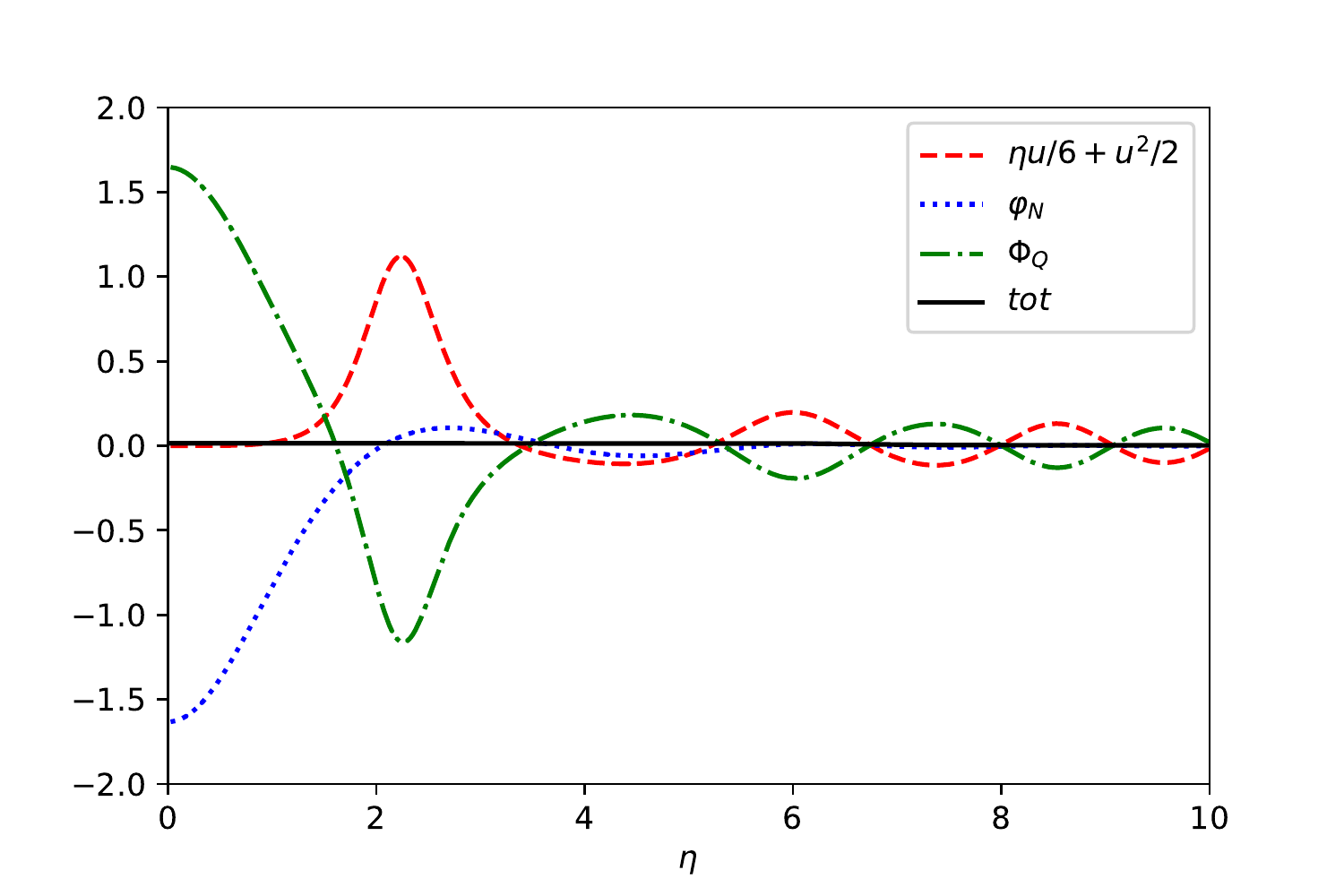}
\caption{{\it Upper panel:} nonlinear density contrast $\hat\delta$ (blue solid line) and linear density contrast $\hat\delta_L$ (red dashed line), for $\hat\delta(0) = 10$.
{\it Middle panel:} nonlinear and linear velocity fields.
{\it Lower panel:} comparison of the terms in the Bernoulli equation (\ref{eq:Bernoulli}).
We show the kinetic part (red dashed line), the Newtonian gravitational potential (blue dotted line), the quantum pressure (green dot-dashed line), and their sum which must be zero (black solid line).}
\label{fig:non-linear-10}
\end{figure}

\begin{figure}[h!]
\centering
\includegraphics[width=0.49\textwidth]{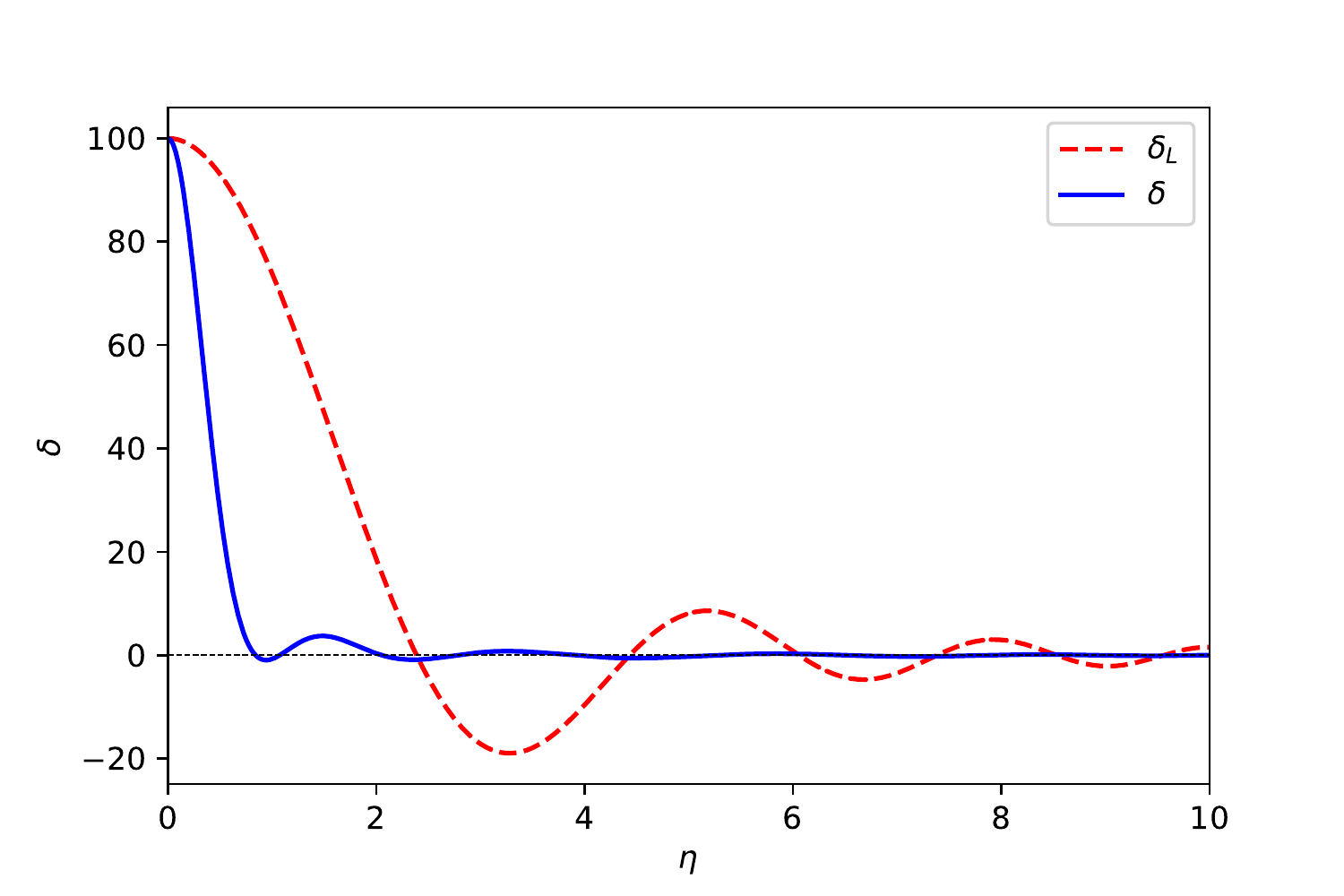}
\includegraphics[width=0.49\textwidth]{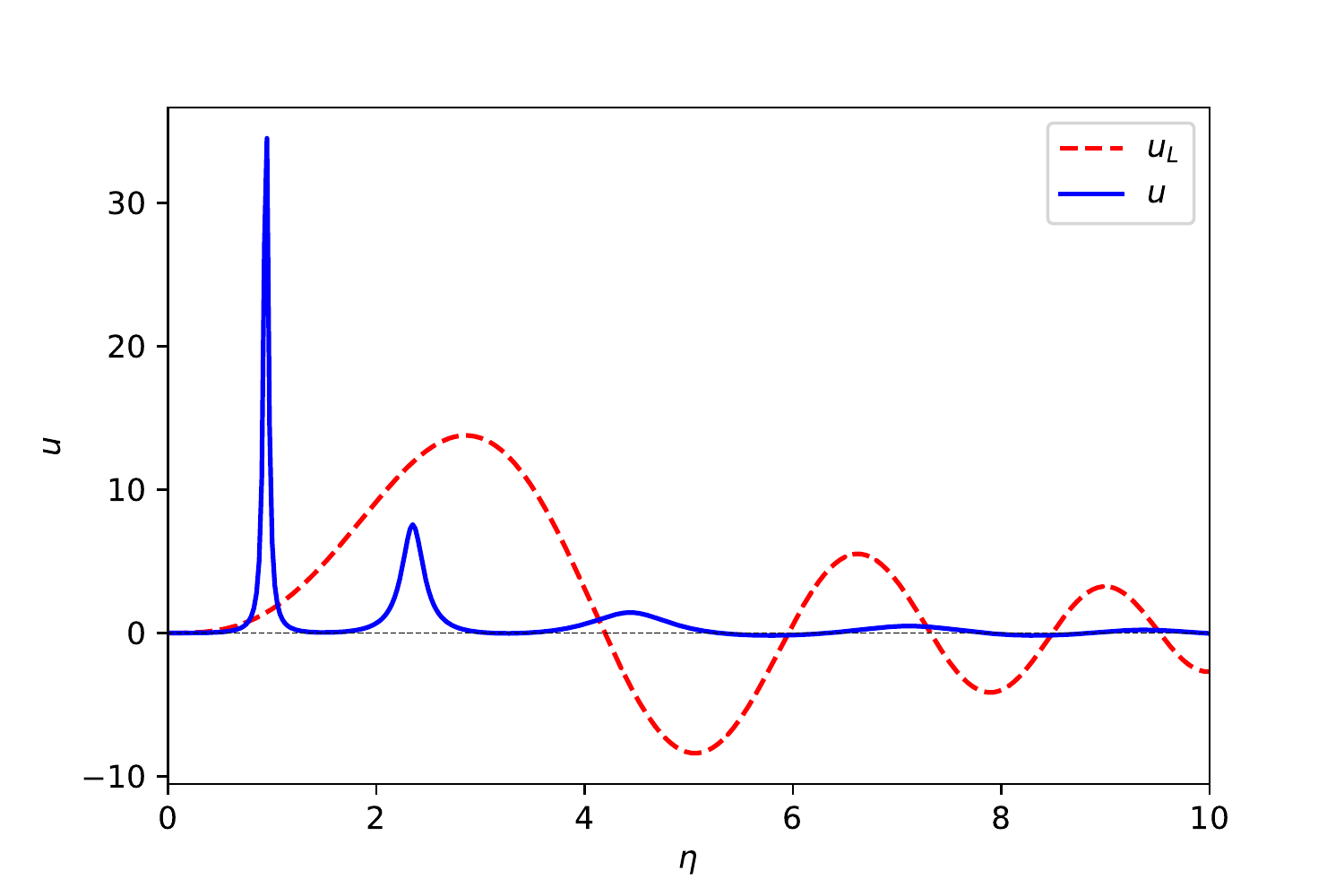}
\includegraphics[width=0.49\textwidth]{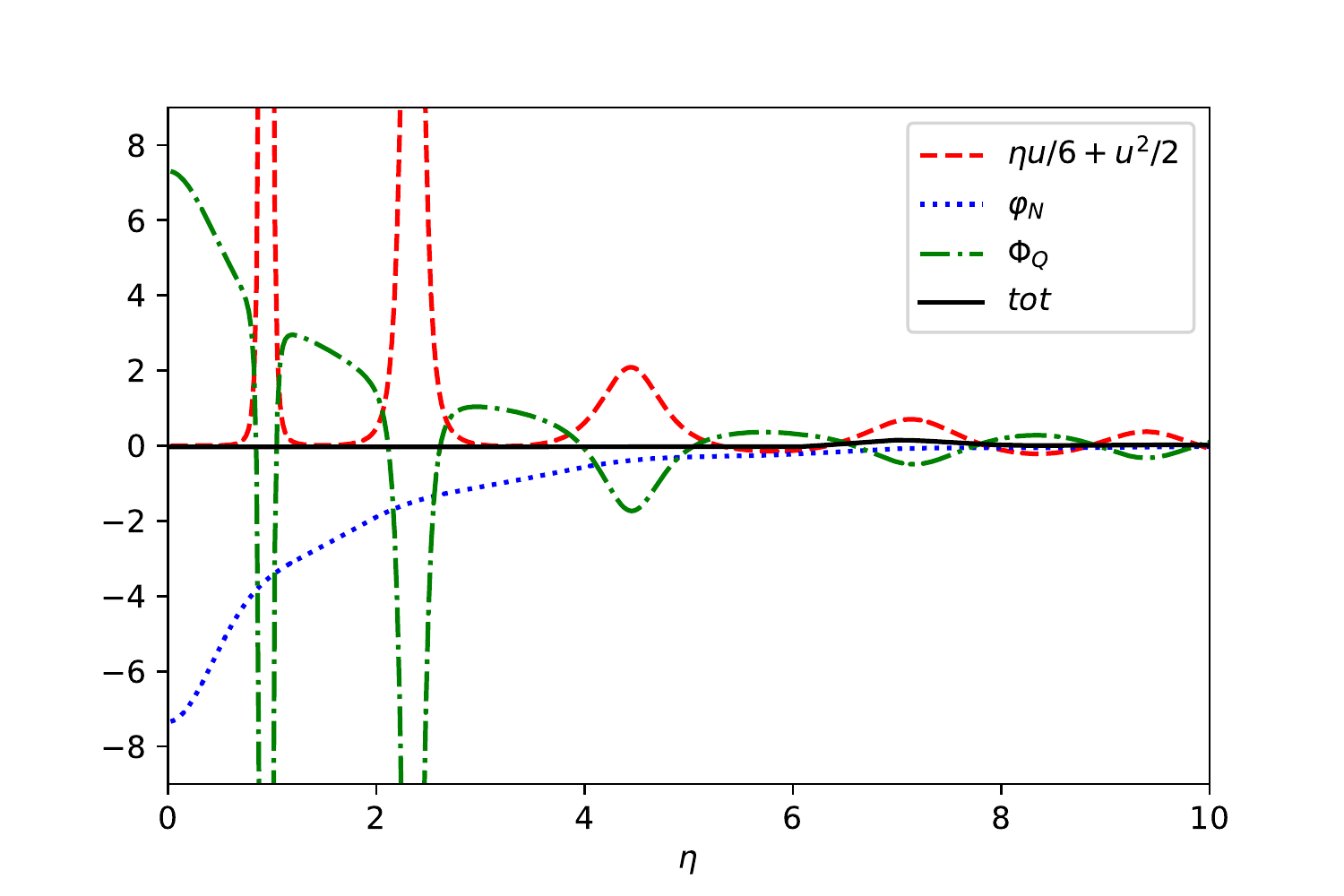}
\caption{Nonlinear and linear solutions as in Fig.~\ref{fig:non-linear-10}, but for
$\hat\delta(0)=100$.}
\label{fig:non-linear-100}
\end{figure}

We compare in Figs.~\ref{fig:non-linear-10} and \ref{fig:non-linear-100} the nonlinear
and linear densities and the velocity perturbations, for the cases
$\hat\delta(0) = 10$ and $100$.
We can see that as the height of the central density peak grows, the nonlinear corrections make
the peak narrower and all the higher-order peaks move closer to the center.
The oscillations of the velocity field also grow and become much sharper. They are no longer
symmetric and the velocity shows high and narrow positive spikes at the density minima, where
$\hat\delta \simeq -1$ and $\rho \simeq 0$.
This can be understood from the scalar matter flux.
As we have explained in Sec.~\ref{sec:numerical-linear},
the profile shrinks in comoving coordinates and loses mass as time grows.
Then, for the scalar mass to escape from the central peak across the  radius $R_1(t)$,
associated with the first minimum of the density, the velocity must be large to ensure a large
flux despite the small local density.
This also holds for the next few peaks where the density minima are also much below the
background density.

This ejection of scalar matter, through successive clumps that escape from the central
density peak and move to infinity, is also seen in numerical simulations and often called
``gravitational cooling'' \citep{Seidel1994,Guzman2006}. 
This allows the system to relax towards equilibrium configurations,
in spite of the absence of dissipative processes, by ejecting extra matter and energy
to infinity.
Of course, in simulations that follow for instance the collisions of DM halos,
this transient process is somewhat chaotic, whereas in our self-similar solutions it is well
ordered and takes place at all times, up to a rescaling of length and mass scales.
However, it is interesting to recover this process in the simple and semi-analytic
self-similar solutions studied in this paper.
We shall come back to this ejection of matter in Sec.~\ref{sec:trajectories} below,
where we compute the trajectories of constant-mass shells.

As explained in Sec.~\ref{sec:numerical-linear} from the linear solution, these dynamics
involve a changing interplay between gravity, quantum pressure and kinetic effects.
This phenomenon remains valid in the nonlinear regime, as seen in the lower panels
that show the various terms in the Bernoulli equation (\ref{eq:Bernoulli}).
Near the center, gravity and the quantum pressure play the major roles,
as the velocity vanishes at the center by symmetry.
Gravity ensures that the central overdensity does not decay too fast and actually follows
the cosmological background density, whereas the quantum pressure resists the gravitational
pull and ejects some matter outside of the central peak.
As the central density becomes very large (see Fig.~\ref{fig:non-linear-100}), this dominance
of gravity and quantum pressure extends over several density peaks, with sharp transitions
at the density minima associated with the spikes of the velocity field, and hence of the
kinetic terms, balanced by spikes of the quantum pressure.
Indeed, being an integral of the matter density, the gravitational potential $\varphi_{\rm N}$ is
very smooth and cannot follow the sharp changes of the velocity. In contrast, the
quantum pressure (\ref{eq:Phi-Q-rho}) being the second derivative of the density, it is a local
quantity that responds to local changes of the system.
At larger distances, where we recover the linear regime, the gravitational potential becomes
negligible as the mass perturbation is screened, like for a compensated profile.
There, the quantum pressure and kinetic terms play the dominant roles and scalar fluctuations
travel afar in a wavelike fashion.
Thus, we can see how the gravitational cooling is related to the wavelike features of
the FDM dynamics, thanks  the relevance  of the quantum pressure.

The numerical computation shows that the density contrast always remains above $-1$,
and hence the density never vanishes and remains strictly positive.
This can be shown analytically as follows.
Let us look for a regular solution with a vanishing density at the point $\eta_0$,
\be
\delta\hat M(\eta) = a_0 + a_1 (\eta-\eta_0) + a_2 (\eta-\eta_0)^2 + \dots ,
\ee
with the zero-density constraint
\be
\hat\delta(\eta_0) = -1 : \;\;\; a_1 = - \frac{2}{3} \eta_0^2 .
\ee
Substituting into the equation of motion (\ref{eq:deltaM-nonlinear}) determines the coefficients
$a_2, a_3, ....$, and gives the solution
\be
\delta\hat M(\eta) = a_0 - \frac{2}{9} (\eta^3-\eta_0^3) ,
\ee
where $a_0$ remains undetermined. We can check that this is indeed a solution of
Eq.(\ref{eq:deltaM-nonlinear}) and it yields $\hat\delta(\eta)=-1$, that is, the constant
zero-density solution $\rho = 0$. Thus, the only regular self-similar solution where the density
vanishes at a point is actually the homogeneous vacuum.
This implies that the nonlinear density profiles shown in Figs.~\ref{fig:non-linear-10} and
\ref{fig:non-linear-100} can never touch the zero-density threshold.
The fact that the density always remains strictly positive means that the Madelung
transformation (\ref{eq:Madelung}) is always well defined and the $\psi$-field and
$\{\rho,\vec v\}$-hydrodynamical pictures are equivalent. Therefore, the self-similar
solutions obtained from (\ref{eq:deltaM-nonlinear}) in terms of $\{\rho,\vec v\}$
give at once the self-similar solutions in terms of $\psi$ and we do not miss any
by working in the hydrodynamical framework.
In fact, because our solutions are spherically symmetric the spatial dimensionnality
is reduced to one and the radial velocity can always be written as the gradient
of a phase $S= \int dr \, v_r$, so that the Schr\"odinger and hydrodynamical
pictures are equivalent.

\subsection{Underdensities}

\begin{figure}[h!]
\centering
\includegraphics[width=0.48\textwidth]{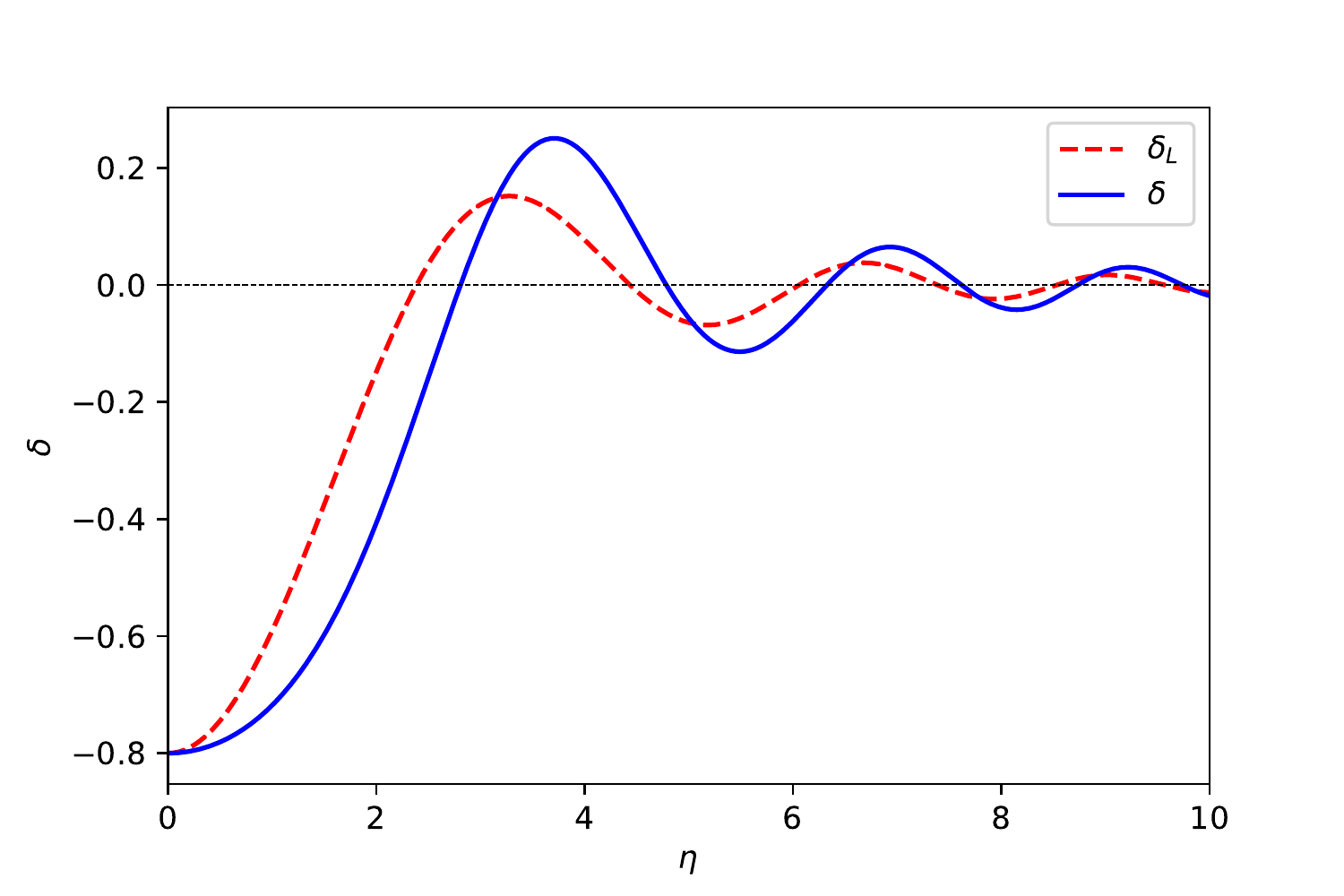}
\includegraphics[width=0.48\textwidth]{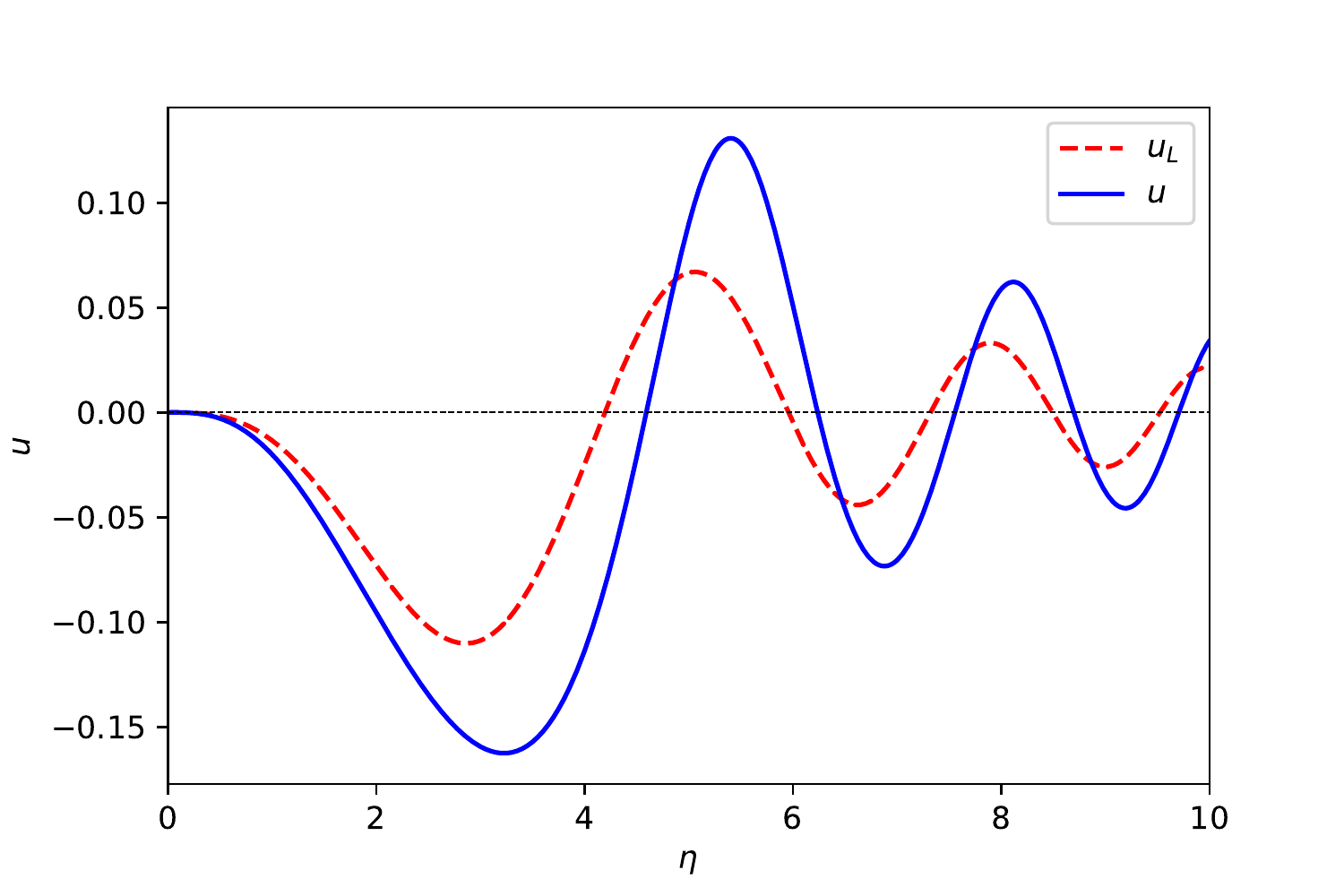}
\includegraphics[width=0.49\textwidth]{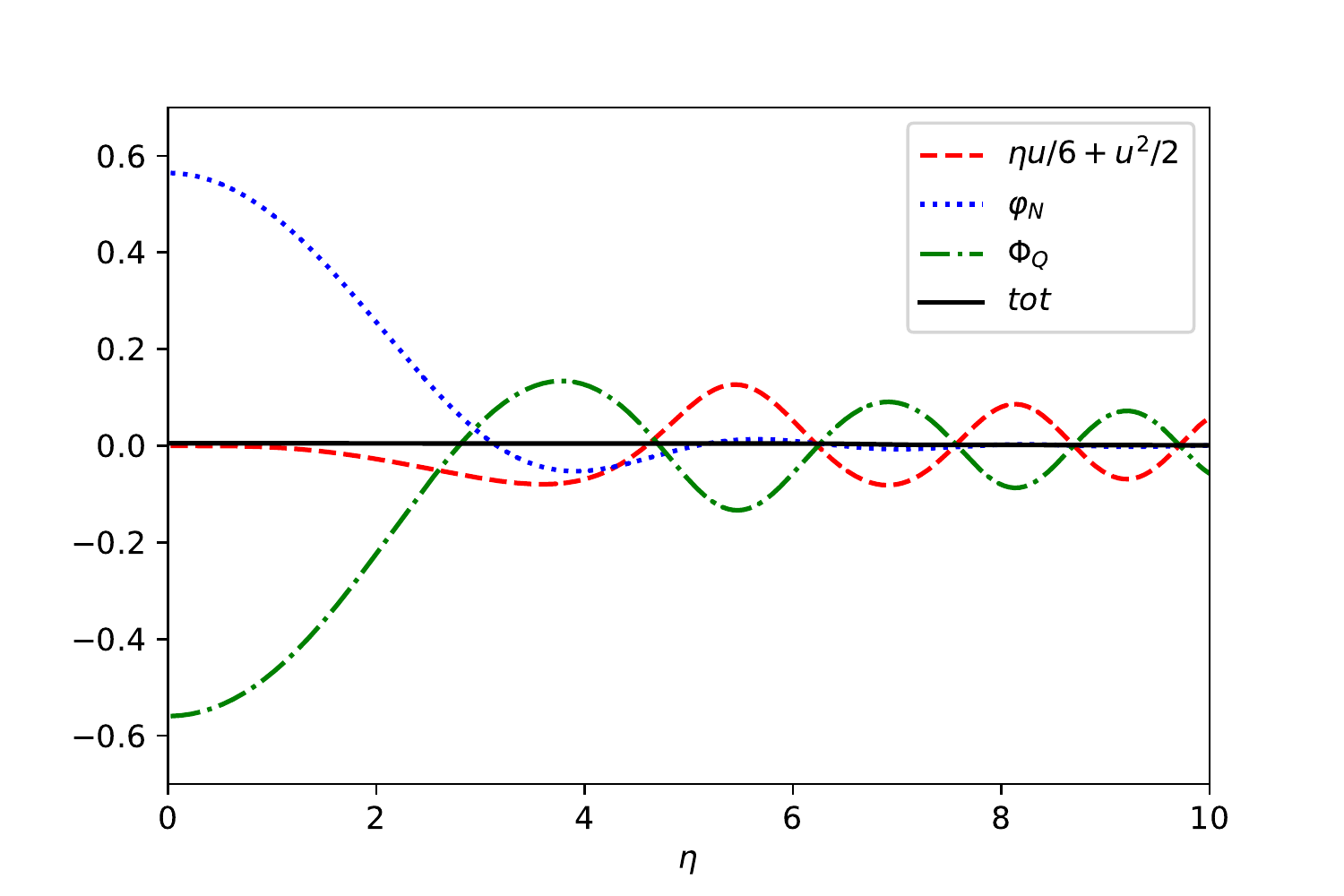}
\caption{Nonlinear and linear solutions as in Fig.~\ref{fig:non-linear-10}, but for
$\hat\delta(0)=-0.8$.}
\label{fig:under-dense-0p8}
\end{figure}

\begin{figure}[h!]
\centering
\includegraphics[width=0.49\textwidth]{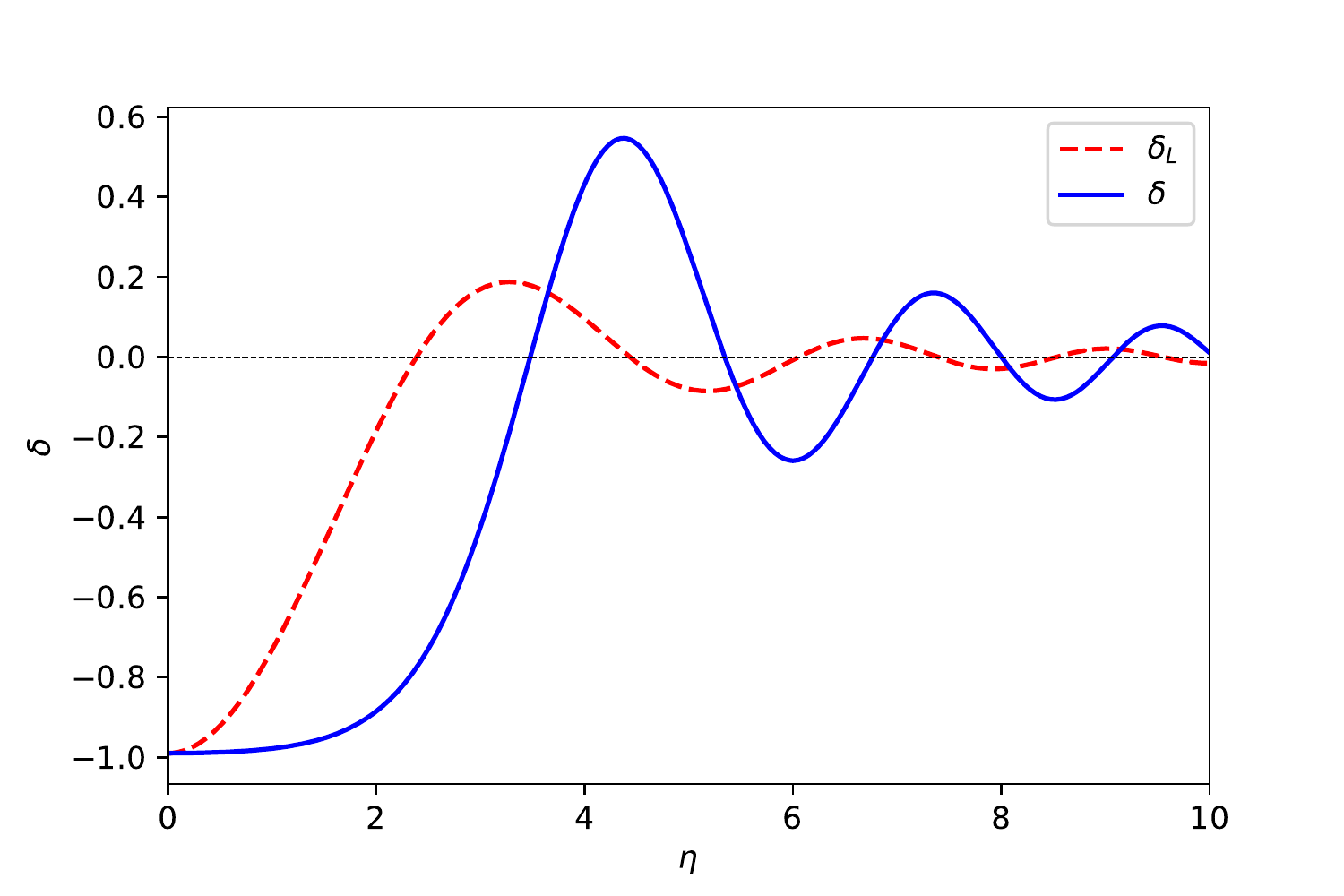}
\includegraphics[width=0.49\textwidth]{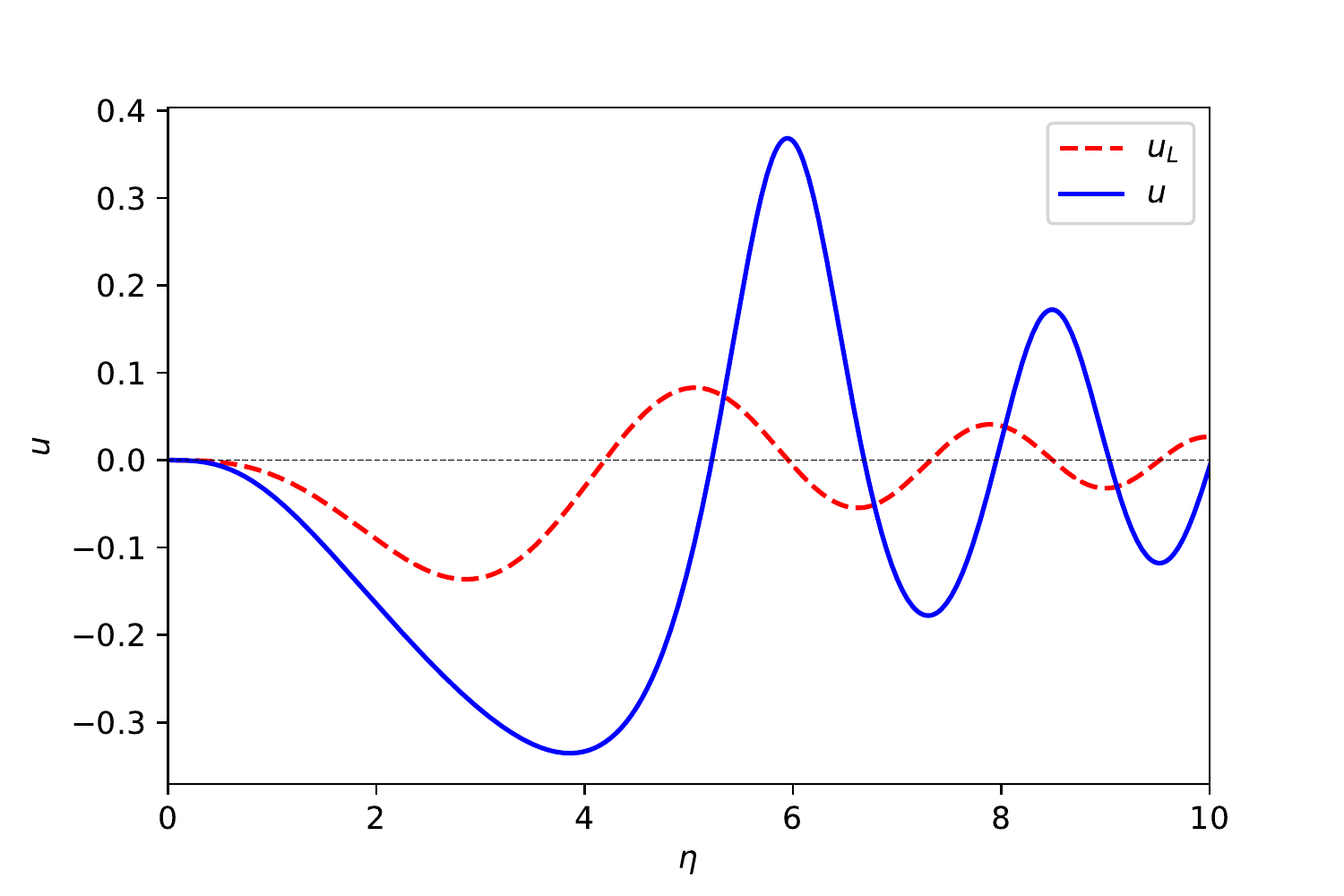}
\includegraphics[width=0.49\textwidth]{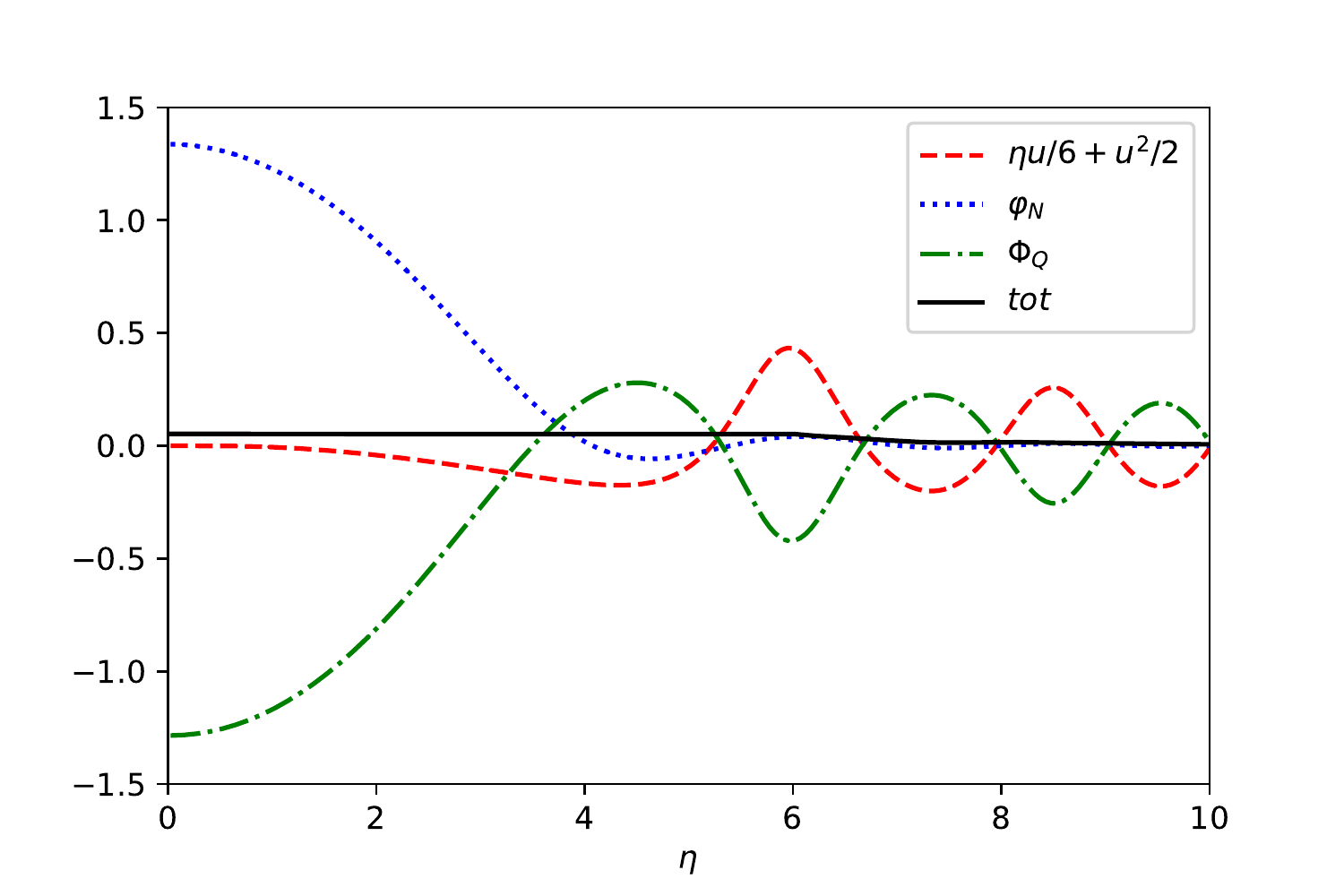}
\caption{Nonlinear and linear solutions as in Fig.~\ref{fig:non-linear-10}, but for
$\hat\delta(0)=-0.99$.}
\label{fig:under-dense-0p99}
\end{figure}

We show in Figs.~\ref{fig:under-dense-0p8} and \ref{fig:under-dense-0p99} the case of underdense
central regions, $\hat\delta(0)=-0.8$ and $-0.99$.
As compared with the linear profiles, we find that nonlinear corrections now move the density
peaks to larger distance and widen the central void, in contrast with the overdense case.
Again, by symmetry the velocity vanishes at the center so that the gravitational and quantum
pressure terms dominate in the Bernoulli equation (\ref{eq:Bernoulli}), while the kinetic
and quantum pressure terms dominate at large distance in the linear regime, where
the mass perturbation is screened as for a compensated profile.

Again, the density contrast always remains above $-1$, that is, the density $\rho$ is always
strictly positive, and the equivalence of the $\psi$-field and
$\{\rho,\vec v\}$-hydrodynamical pictures is valid.

\subsection{Husimi distribution}
\label{sec:num-Husimi}

\subsubsection{Overdensities}
\label{sec:Husimi-overdensities}

\begin{figure*}[ht]
\centering
\includegraphics[width=0.49\textwidth]{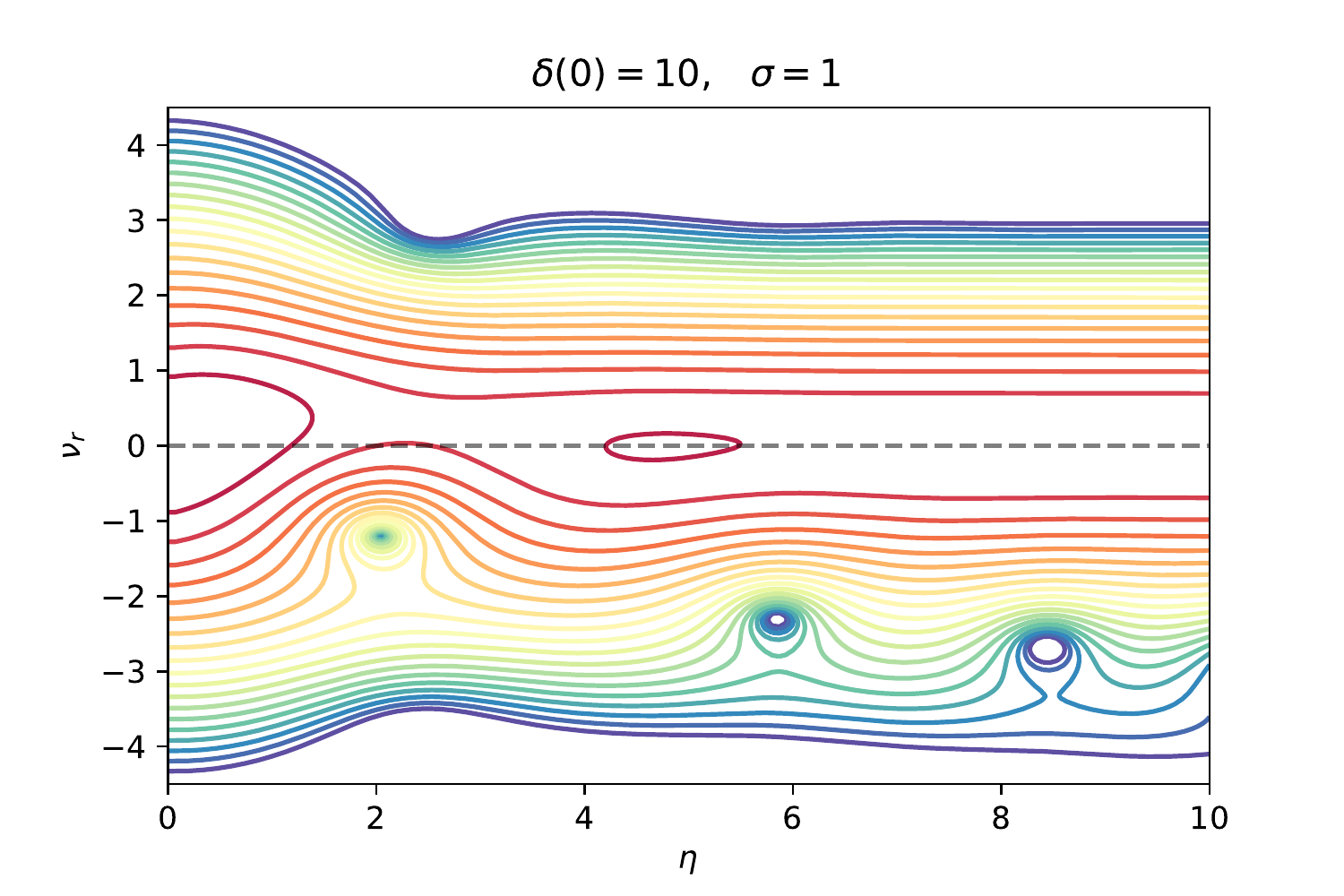}
\includegraphics[width=0.49\textwidth]{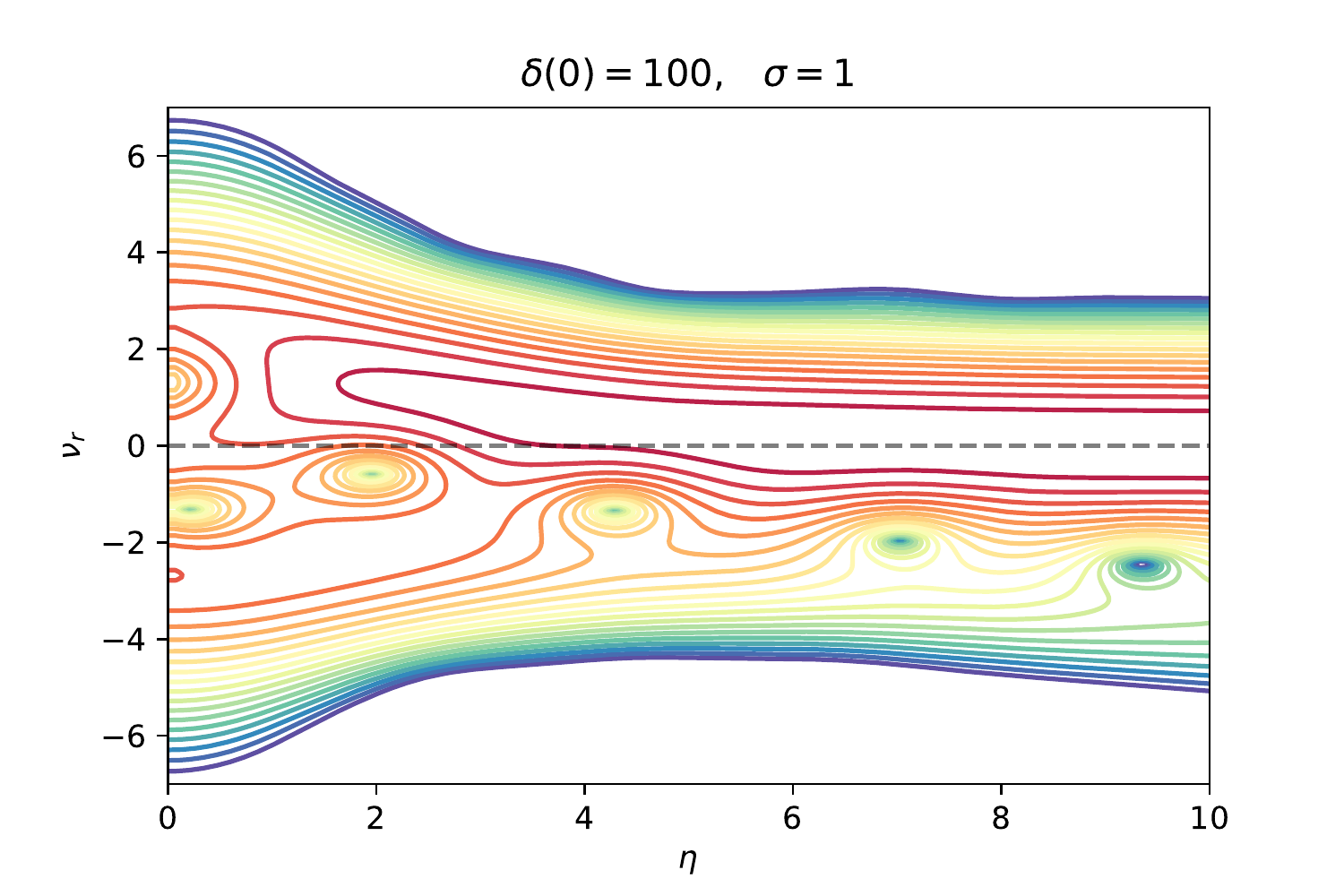}\hfill
\includegraphics[width=0.49\textwidth]{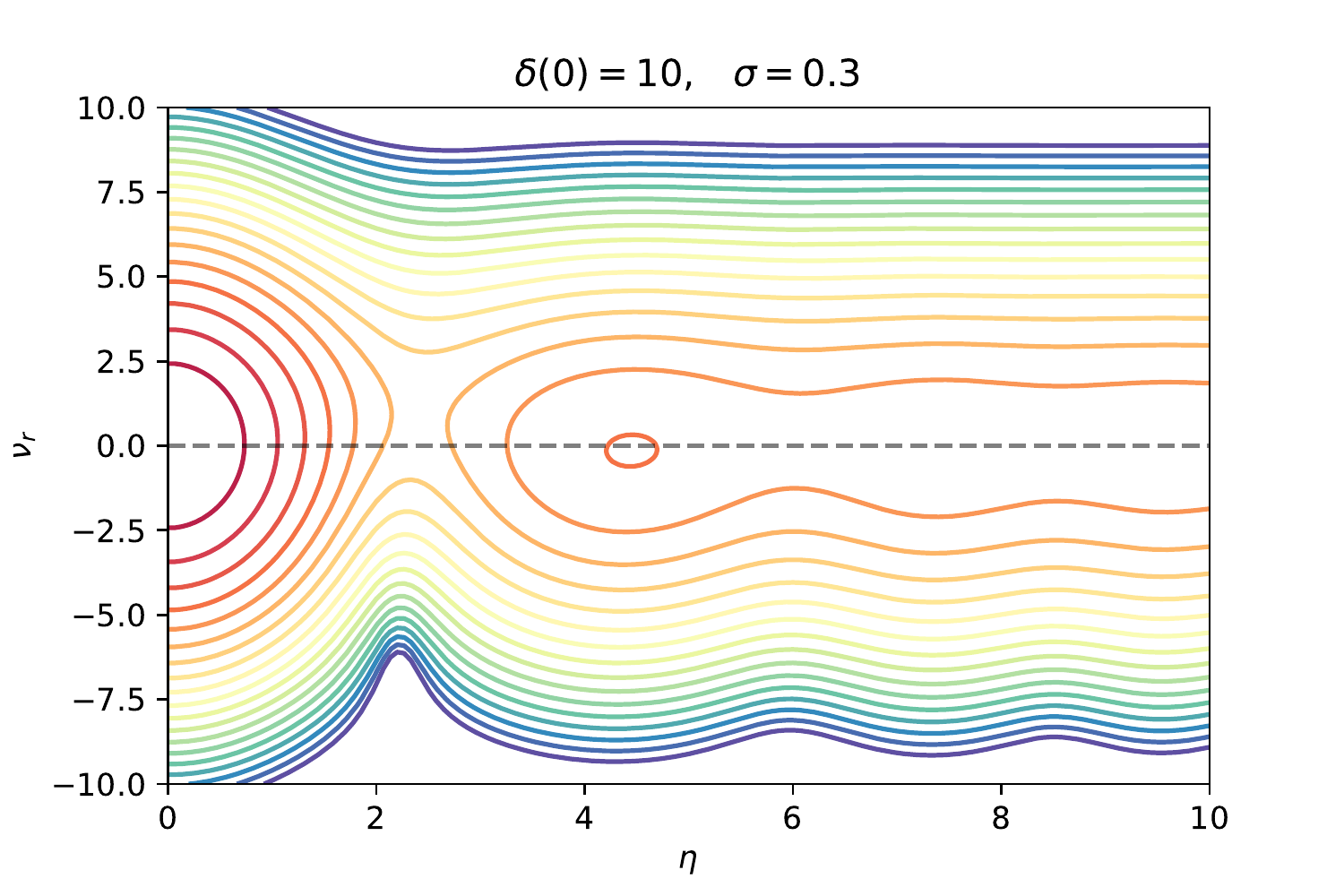}
\includegraphics[width=0.49\textwidth]{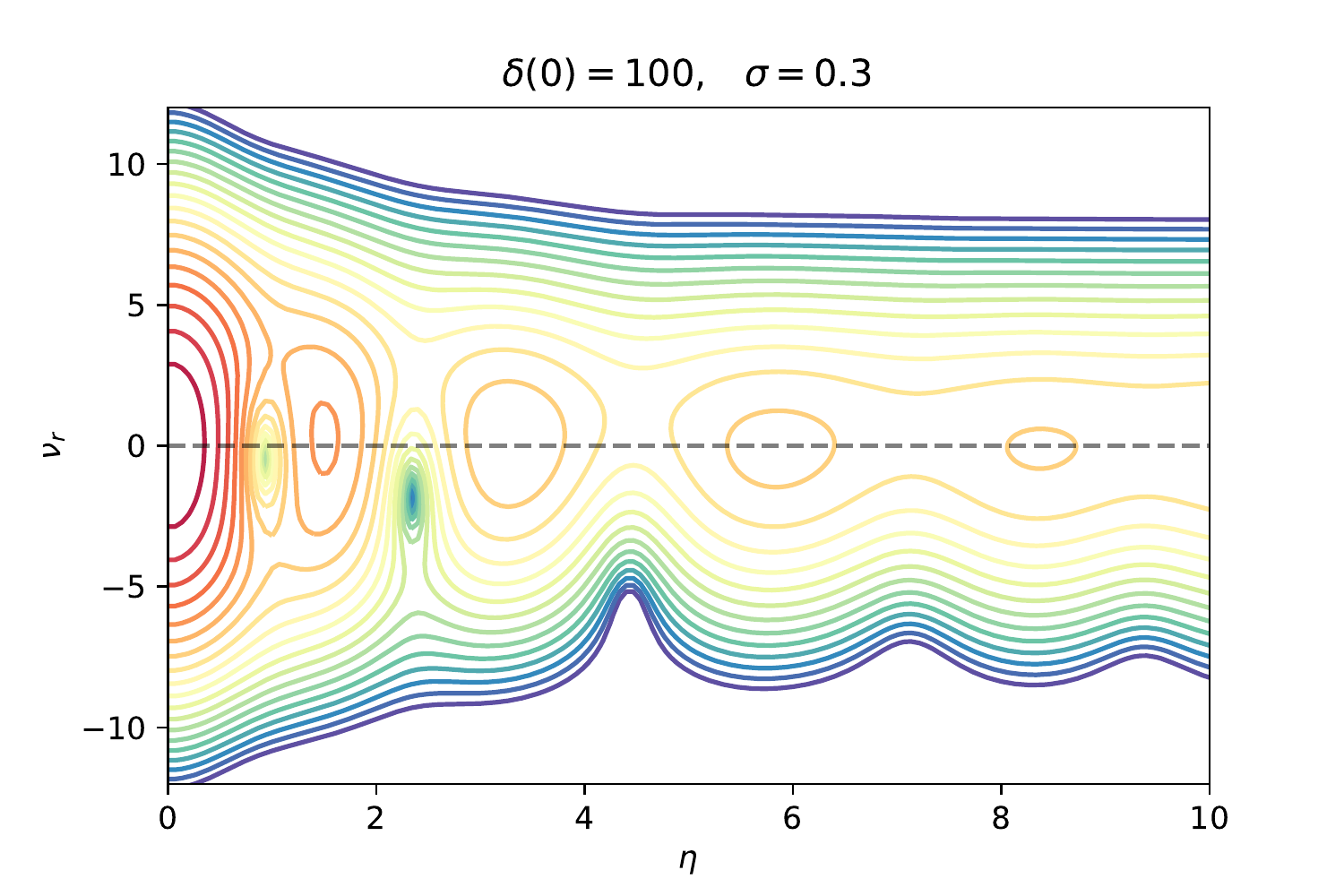}
\caption{Isodensity contours for the radial Husimi distribution function $\hat f_{\rm H}(\eta,\nu_r)$
for $\sigma=1$ (upper row, $\delta(0)= 10$ and $100$) and for $\sigma = 0.3$
(lower row, $\delta(0)= 10$ and $100$).}
\label{fig:husimi-10-100}
\end{figure*}

We show in Figs.~\ref{fig:husimi-10-100} the radial Husimi distributions $\hat f_{\rm H}(\eta,\nu_r)$,
where $\nu_r$ is the radial velocity.
As seen in Figs.~\ref{fig:non-linear-10} and \ref{fig:non-linear-100}, as the central density
increases successive density peaks are more clearly defined and separated by almost void
regions. This is also seen in the Husimi distribution, with the number of well-defined
peaks increasing with $\delta(0)$. At large distance, where the profile converges to the
cosmological background, the finite smoothing $\sigma$ involved in the definition of the
Husimi distribution smears out the density and velocity perturbations and the Husimi
distribution converges to the background result (\ref{eq:fH-background}).

As the spatial coarsening $\sigma_x \propto \sigma$ decreases, the velocity coarsening
$\sigma_p\propto 1/\sigma$ increases, following the Heisenberg uncertainty principle
(\ref{eq:sigma-x-sigma-p}). Thus, for the case $\delta(0)= 10$ with $\sigma=1$,
we can see the velocity asymmetry inside the central peak (the radial velocity is positive as
seen in Fig.~\ref{fig:non-linear-10}) but the spatial profile is somewhat smoothed out.
Decreasing $\sigma$ to $0.3$ gives a better separation between the first two peaks and
preserves signs of the density fluctuations at larger radii, but this comes at the price
of a significant smoothing along the velocity axis and the asymmetry of the velocity distribution
in the central peak can no longer be distinguished.
Note that the scale of the vertical velocity axis is larger for the lower panels with
$\sigma=0.3$.
In fact, in the limit $\sigma\to 0$ we have
$\hat f_{\rm H} \sim \sigma^3 \pi^{-3/2} \hat\rho(\eta)$ at fixed $\vec \nu$.

For $\delta(0)= 100$ the spatial width of the central and few subsequent peaks shrinks,
as seen in Fig.~\ref{fig:non-linear-100}. This implies that the coarsening $\sigma=1$
is no longer sufficient to separate the first few peaks. This leads to artificial interferences
between these peaks and to a Husimi distribution that is difficult to interpret and far from
the semiclassical expectations. Decreasing $\sigma$ to $0.3$ gives a more faithful representation
of the system, as we can clearly see the sequence of scalar-field clumps. However, this
erases most of the information about the velocity field.

This shows that it is not always straightforward to use the Schr\"odinger equation
(\ref{eq:psi-eq-comoving}) and the Husimi distribution as an alternative to N-body simulations
to compute the classical phase-space distribution that obeys the Vlasov equation.
Different choices of the smoothing $\sigma$ can lead to rather different pictures that can be
difficult to relate to the underlying dynamics.
This may become a problem for systems that exhibit a large range of scales, as for the
hierarchical gravitational clustering displayed by cosmological structures.
For the self-similar solutions that we study in this paper, where the density is everywhere
strictly positive and the hydrodynamical mapping (\ref{eq:Madelung}) well defined,
the density and velocity fields provide a clearer picture of the dynamics than the
phase-space Husimi distribution.

\subsubsection{Underdensities}
\label{sec:Husimi-underdensities}

\begin{figure}[h!]
\centering
\includegraphics[width=0.49\textwidth]{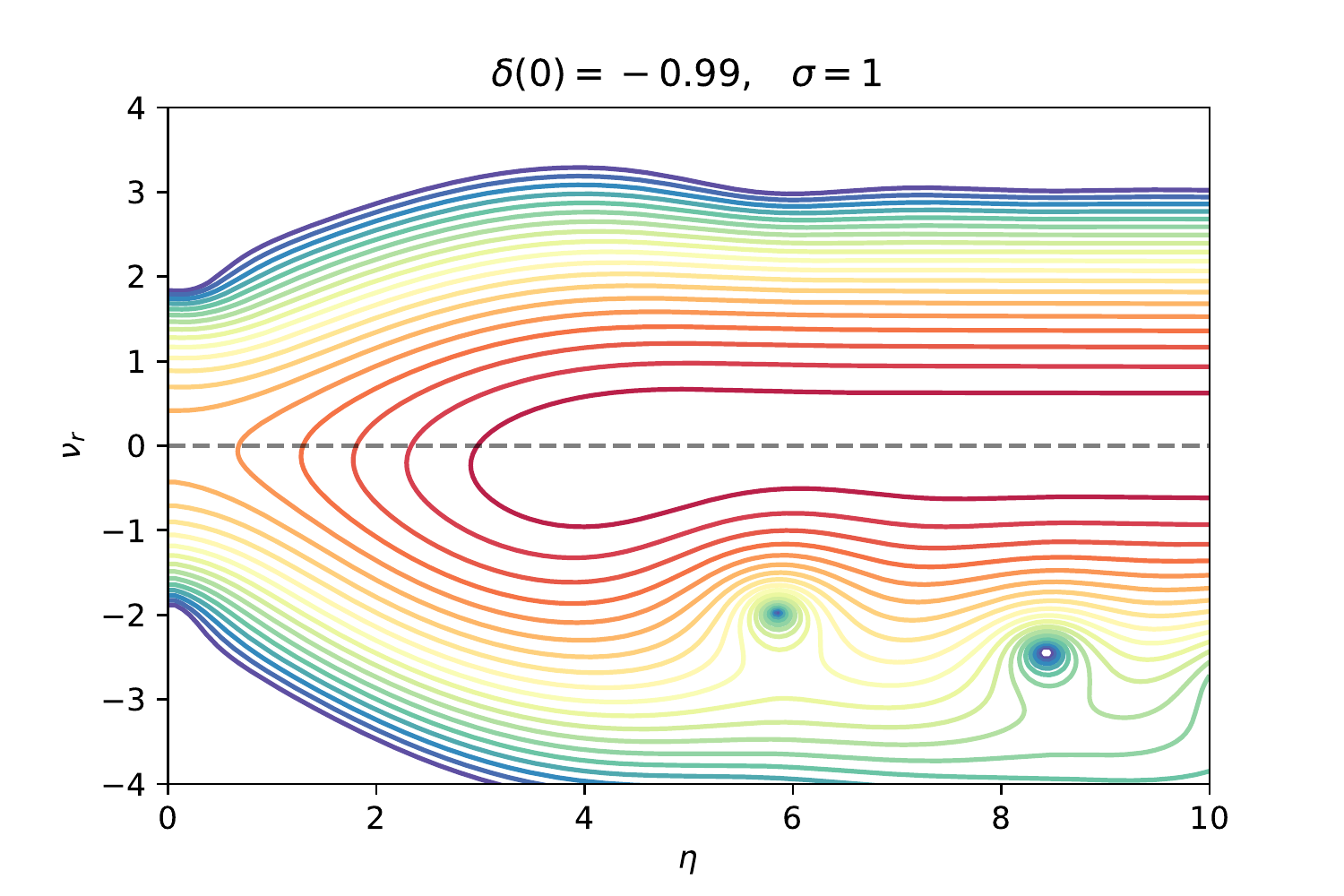}
\includegraphics[width=0.49\textwidth]{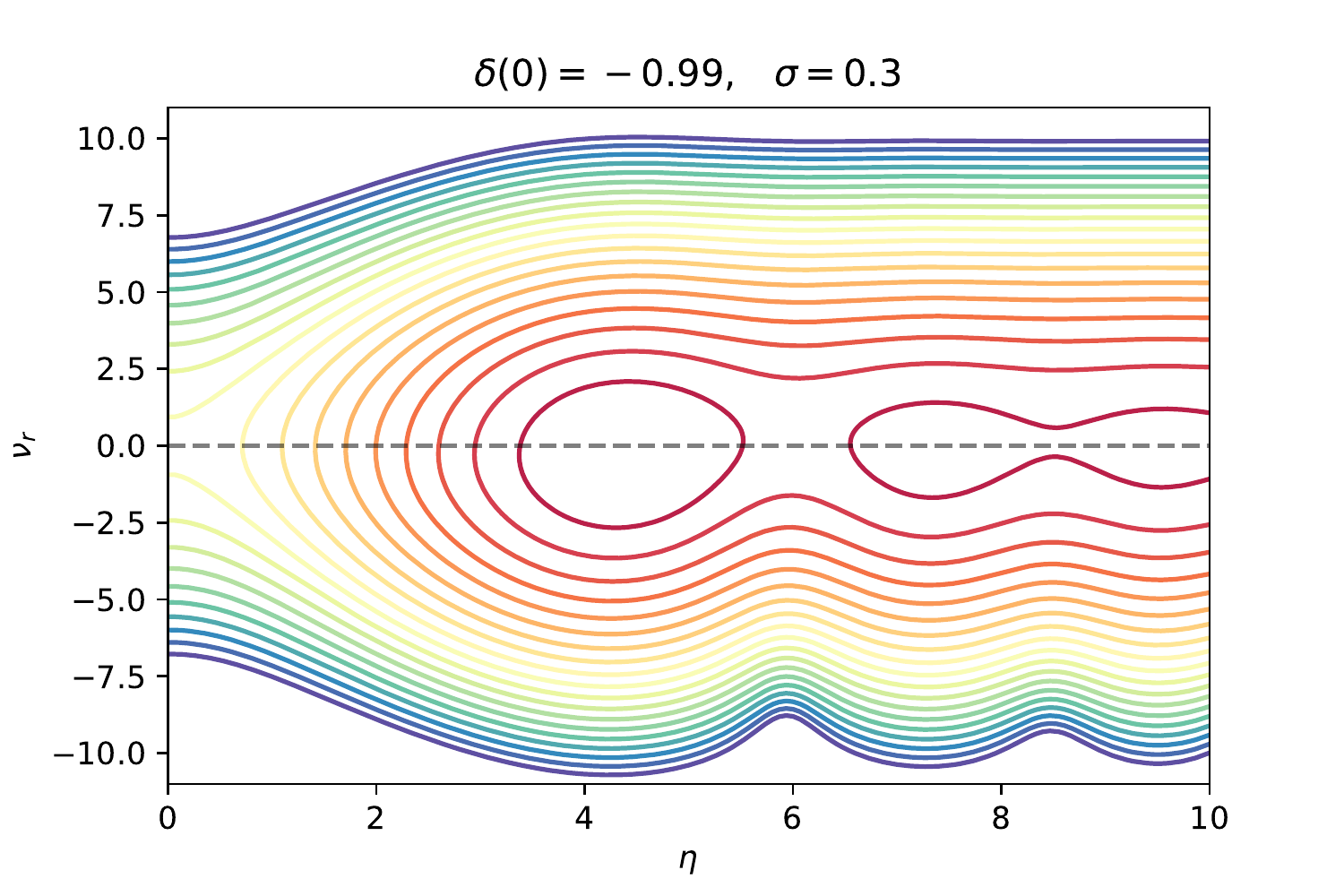}
\caption{Isodensity contours for the radial Husimi distribution function $\hat f_{\rm H}(\eta,\nu_r)$,
for $\sigma = 1$ (upper panel) and $\sigma=0.3$ (lower panel) with $\delta(0)= -0.99$.}
 \label{fig:husimi-0p99}
 \end{figure}

We show in Fig.~\ref{fig:husimi-0p99} the case of the central underdensity $\delta(0)=-0.99$.
Again, we find that the smaller value of $\sigma$ enables a better separation of the successive
density peaks but erases the velocity asymmetries.

\subsection{Trajectories associated to the self-similar solutions}
\label{sec:trajectories}

\begin{figure}[h!]
\centering
\includegraphics[width=0.49\textwidth]{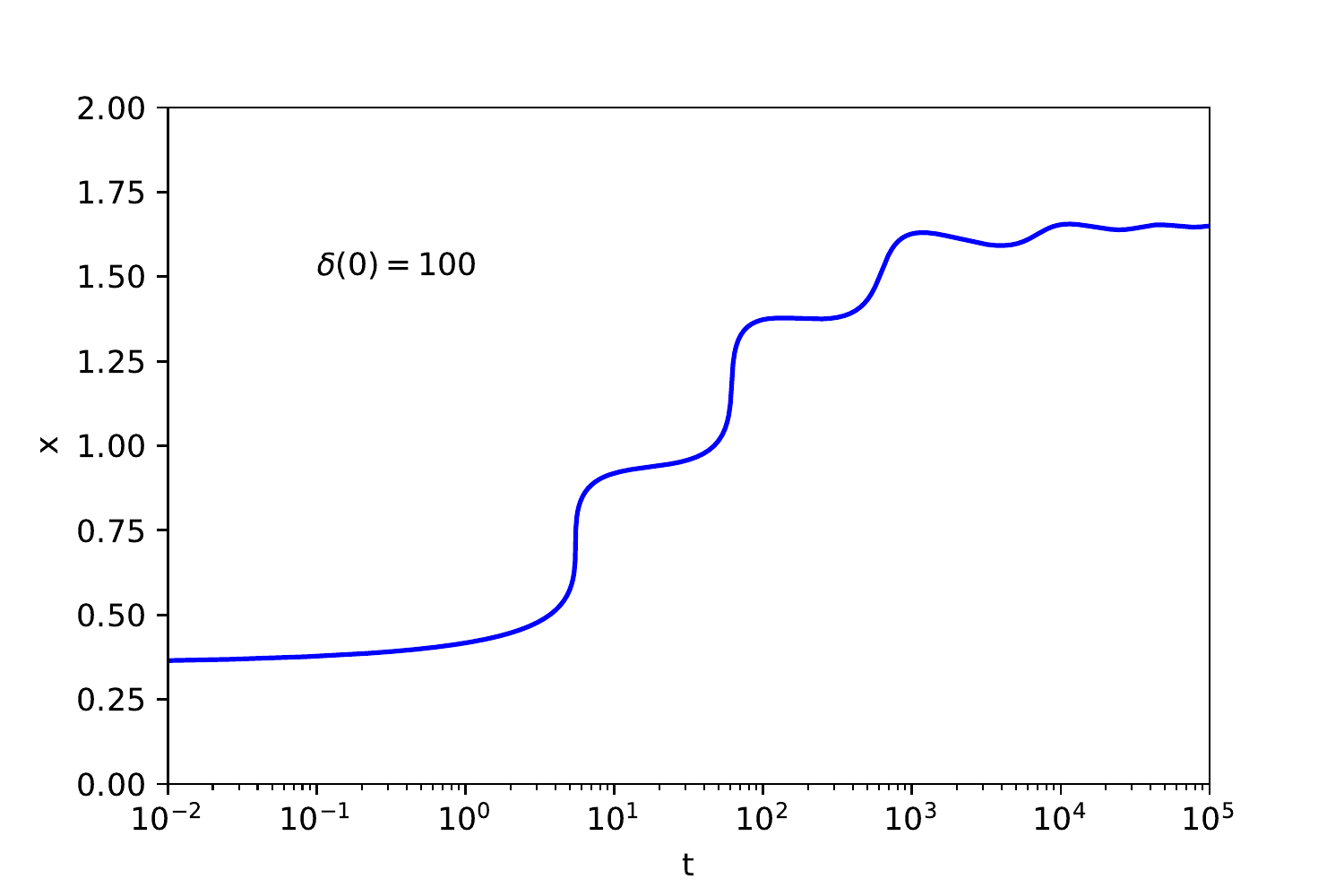}
\includegraphics[width=0.49\textwidth]{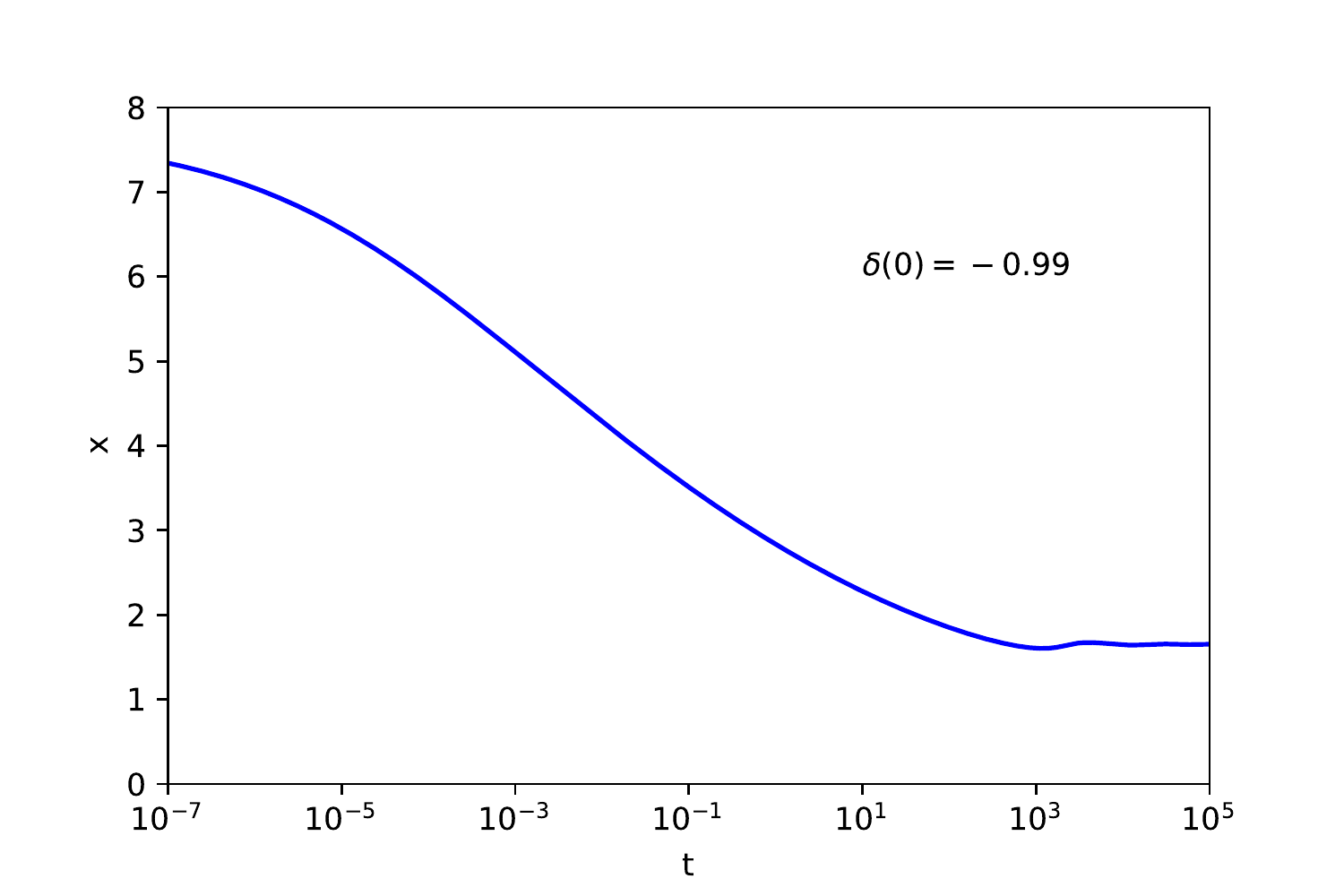}
\caption{Trajectory $x(t)$ of the comoving radius associated with a fixed mass, as a function
of cosmic time $t$. We show the case of the self-similar solutions defined by the central
density contrasts $\delta(0)=100$ (upper panel) and $\delta(0)=-0.99$ (lower panel).}
 \label{fig:trajectory}
 \end{figure}

The density and velocity fields, as well as the Husimi distribution, provide an Eulerian
point of view on the self-similar solutions. To  understand better the dynamics it is useful
to consider a complementary Lagrangian point of view. Thanks to the spherical symmetry,
in the hydrodynamical picture the analogue of a particle trajectory is the motion of the radius $r(t)$
that encloses a fixed mass $M(<r)=M$.
From Eq.(\ref{eq:self-r-x}), the mass $M(<r) = \bar M + \delta M$ reads
\be
M = \bar{M}+ \delta M = \epsilon^{3/2}t^{-1/2}\left[ \frac{2}{9}\eta^3+\delta\hat M(\eta) \right] .
\label{eq:mass-shell}
\ee
This implicitly gives the trajectory $\eta(t)$ as a function of time for a given mass.
In fact, from Eq.(\ref{eq:mass-shell}) we obtain at once the scaling law
\be
\eta(t | M,\epsilon) = \eta( \epsilon^{-3} M^2 t ) ,
\label{eq:eta-M-t}
\ee
that is, $\eta$ only depends on the combination $\epsilon^{-3} M^2 t$.
Therefore, trajectories associated with different masses or different values of $\epsilon$
can be obtained from a single trajectory by a rescaling of time,
\be
t \propto \epsilon^3 M^{-2} .
\label{eq:t-scaling}
\ee
However, the shape of the trajectory depends on the self-similar profile, defined for instance
by the central density contrast $\delta(0)$.
The result (\ref{eq:t-scaling}) shows that large masses correspond to small times, in contrast
with the self-similar CDM solutions that describe a hierarchical collapse where large masses
collapse later \cite{Fillmore:1984wk,Bertschinger:1985pd}.
This is because the FDM self-similar solutions describe instead a slow blow-up, that roughly
follows the Hubble expansion.
Indeed, as the total overdensity $1+\delta$ is always positive, at fixed time $M(\eta)$
and $\eta(M)$ are monotonic increasing functions. The scaling law (\ref{eq:eta-M-t})
then implies that $\eta(t)$ is also a monotonic increasing function of $t$ at fixed $M$.
Therefore, at $t\to 0$ we have $\eta \to 0$ and at $t\to\infty$ we have $\eta \to \infty$.
Thus, in terms of the rescaled radius $\eta$, at early time the mass shell starts close to origin,
inside the central peak or void, far in the nonlinear regime if the self-similar solution is nonlinear
at the center, whereas at late time the mass shell moves increasingly far in the linear regime,
at large distance.
Thus, whereas the trajectories found in the CDM self-similar solutions describe a spherical
collapse that runs from the linear to the nonlinear regime, the trajectories found in the
FDM self-similar solutions describe an expansion that runs from the nonlinear to the linear regime,
independently of whether the central region is overdense or underdense.

We show in Fig.~\ref{fig:trajectory} the trajectories obtained inside the self-similar solutions
defined by $\delta(0)=100$ and $-0.99$. We plot the trajectories $x(t)$ in terms of the
comoving coordinate, using Eq.(\ref{eq:eta-def-x}), which gives $x(t) = \epsilon^{1/2} t^{-1/6} \eta(t)$.
For the numerical computations we take $\epsilon=1$, $M=1$; as explained in
Eq.(\ref{eq:t-scaling}) other values of $\epsilon$ or $M$ only correspond to a rescaling of time
and radius.
We can see that in all cases the trajectories roughly follow the Hubble expansion, as the
comoving radius $x(t)$ goes to nonzero finite values at both small and large times.
This agrees with the scaling $\delta=\hat\delta(\eta)$ in (\ref{eq:self-r-x}),
which means that the typical density $\rho$ follows the decrease of the background density
$\bar\rho \propto t^{-2}$. Indeed, this clearly implies that mass shells cannot expand much more
slowly or faster than the Hubble flow.
At late times, which correspond to large $\eta$, the background term
$2\eta^3/9$ dominates in the bracket in Eq.(\ref{eq:mass-shell}). This means that,
independently of $\delta(0)$, we recover the Hubble flow with $x(t) \simeq \bar x$,
where $\bar x = (9/2)^{1/3} M^{1/3}$ is the background comoving radius associated with
the mass $M$. On top of this asymptotic value we have subdominant oscillations associated
with the linear regime.
This also corresponds to the limit of large masses, as they are at larger radii further into
the linear regime.

For the overdense case, we can see that the comoving radius $x(t)$ increases from its initial
to its final value. This is because the mass shell is initially inside the central density peak,
close to the origin. This overdense configuration implies an initial radius $x_i$ that is smaller
than its counterpart $\bar x$ in the background universe, for the same mass $M$
(as $\rho \propto M/x^3$). At late time, as we recover the Hubble flow with increasingly
small perturbations, the trajectory converges to $\bar x > x_i$.
Conversely, for the underdense case, the comoving radius $x(t)$ decreases from its initial
to its final value.
In the case of the high central overdensity $\delta(0)=100$, where there are three well
marked density peaks separated by velocity spikes, see Fig.~\ref{fig:non-linear-100},
the trajectory has an intermittent character with well-distinguishable steps in the nonlinear regime.
The comoving radius grows very slowly when the shell is inside the density peaks,
or clumps, and shows fast accelerations as it moves from one clump to the next, because of the
velocity spikes found in Fig.~\ref{fig:non-linear-100}, associated with the voids that
separate the clumps.
For the underdense case $\delta(0)=-0.99$ the secondary peaks and their velocity spikes
are weak, see Fig.~\ref{fig:under-dense-0p99}, and we cannot distinguish well-marked steps
in the trajectory.

Although the self-similar profile shrinks in comoving coordinates, as $x \propto t^{-1/6}$
at fixed $\eta$ from Eq.(\ref{eq:eta-def-x}), the mass-shell trajectories remain roughly constant,
with a global finite increase for overdense cases and a global finite decrease for underdense
cases.
Thus, as for wave packets where one needs to distinguish the group and phase velocities,
we can distinguish two velocities or trajectories in the self-similar solutions.
We have a ``geometric'' trajectory, $x \propto t^{-1/6}$, that describes the shrinking
of the self-similar profile, and a ``matter'' trajectory, $x \sim {\rm constant}$, that describes
the flow of matter. They correspond to $r \propto t^{1/2}$ and $r\sim t^{2/3}$ in physical
coordinates.
In particular, as explained above, matter flows ``through'' the self-similar profile towards
the large-distance linear regime. It escapes from the nonlinear central region, going
through a series of clumps and velocity bursts, until it converges to the Hubble flow at large
distance.
This ejection of matter is reminiscent of the gravitational cooling found in numerical
simulations.

Therefore, the matter content of a given clump does not remain fixed with time,
as matter slowly flows through it (and reaches large velocities at its boundaries, where the density
is very small).
Again, this is reminiscent of the behaviors of system governed by wave equations.
Here this is due to the quantum pressure, which originates from the Schr\"odinger
equation that exhibits well-known wavelike and interference behaviors.

\section{High-density asymptotic limit}
\label{sec:asymptotic}

\begin{figure}[h!]
\centering
\includegraphics[width=0.49\textwidth]{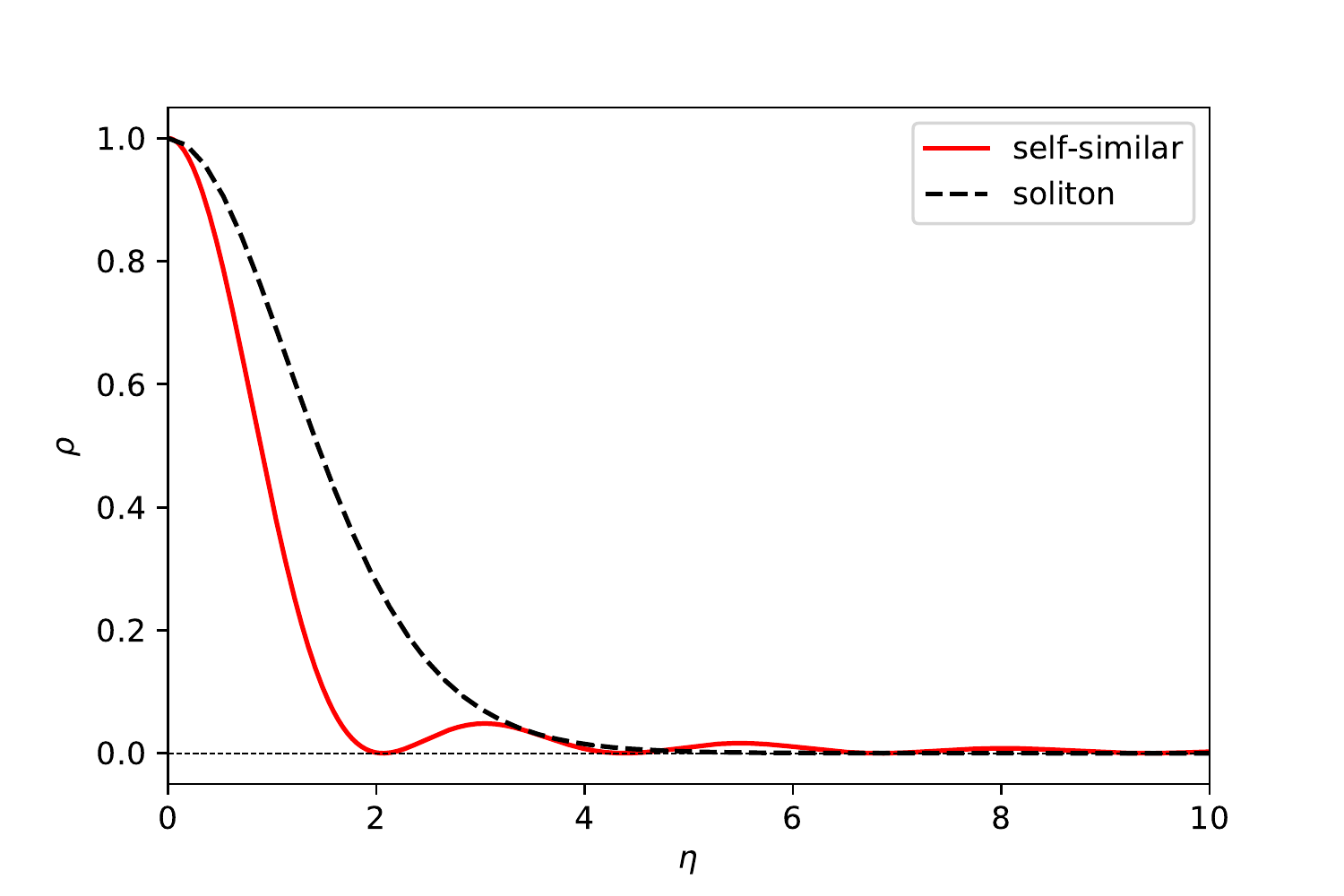}
\caption{Asymptotic self-similar (red solid line) and soliton (black dashed line) density profiles,
normalized to $\rho(0)=1$.}
 \label{fig:asymptotic-1}
 \end{figure}

We recalled in Sec.~\ref{sec:eqs-of-motion} that the SP system
(\ref {eq:Schrodinger-real-1})-(\ref{eq:Poisson-real-1}) is invariant under the scaling law
(\ref{eq:res1}). In the case of the cosmological self-similar solutions, this symmetry is broken by
the Einstein-de Sitter background that sets the boundary condition at large distance.
In contrast with the family of solitonic solutions (\ref{eq:soliton-eq}) in vacuum,
we can no longer obtain a family of self-similar solutions by the rescaling (\ref{eq:res1}),
as this would also change the density at infinity (and hence correspond to a different boundary
condition).
However, in the limit of large density contrasts, the background density becomes negligible
as compared with the central density and we can expect the inner profile to converge
to a limiting shape that obeys the scaling law (\ref{eq:res1}), which reads here 
\be
\left\lbrace \eta, \psi , \rho , M \right\rbrace \rightarrow \left\lbrace \lambda^{-1} \eta , \lambda^{2}\psi , \lambda^4 \rho , \lambda M \right\rbrace .
\label{eq:res2}
\ee
We can obtain the equation satisfied by this limiting profile by substituting these scalings
into Eq.(\ref{eq:deltaM-nonlinear}) and keeping only the leading terms in the limit
$\lambda\to\infty$.
This gives
\be
M'^2 M^{(4)} - 2 M' M'' M^{(3)} + M''^3 = \frac{4}{\eta^2} M M'^3 \; .
\label{eq:asymptotic}
\ee
In this limit we identified $M\simeq \delta \hat M$ and $\rho \simeq \hat\delta$, so that the
density is given by $\rho = 3 M'/(2\eta^2)$.
This nonlinear equation is then invariant under the symmetry (\ref{eq:res2}).
This means that a complete family of solutions can be obtained from one solution,
normalized for instance by $\rho(0)=1$, by the rescaling (\ref{eq:res2}).
Solving Eq.(\ref{eq:asymptotic}) is actually more difficult than finding solutions of
 Eq.(\ref{eq:deltaM-nonlinear}), as the density minima between the successive
peaks now touch the vacuum value $\rho=0$. These points, where $M'=M''=0$,
are singular points of the differential equation (\ref{eq:asymptotic}).
In practice, we compute the finite-$\lambda$ profile defined by Eq.(\ref{eq:deltaM-nonlinear})
and check that for large $\delta(0)$ the curves collapse to a unique profile
normalized to $\rho(0)=1$ after applying the scaling (\ref{eq:res2}).
We also checked that this profile approximately satisfies Eq.(\ref{eq:asymptotic}).

We compare in Fig.~\ref{fig:asymptotic-1} this asymptotic self-similar profile with the soliton
profile obtained from Eq.(\ref{eq:soliton-eq}), which reads in terms of the dimensionless
variables
\be
\nabla^2_\eta \psi_{\rm sol} = 2 (\varphi_{\rm N}-\alpha) \psi_{\rm sol} , \;\;\;
\nabla^2_\eta\varphi_{\rm N} = \frac{2}{3} \psi_{\rm sol}^2  .
\ee
We can see that the two profiles do not coincide. Thus, even as the central density increases the
shape of the central peak of the self-similar profile does not converge to the
soliton equilibrium. This is due to the importance of kinetic effects, which dominate near
the boundary of the central peak as seen in the lower panel in Fig.~\ref{fig:non-linear-100}.
Besides, beyond the kinetic terms, the soliton balance equation (\ref{eq:hydrostatic}) differs
from the self-similar Bernoulli equation (\ref{eq:Bernoulli}) by the $\alpha$ parameter on the
right-hand side. Thus, the two profiles are different and the central peak of the self-similar
solution is narrower than the soliton peak.
Therefore, even though at large densities the local timescale inside the central peak becomes
much smaller than the Hubble time, the profile does not relax towards the static soliton profile.
This shows that the convergence towards the soliton core is not guaranteed in all configurations.

\begin{figure}[h!]
\centering
\includegraphics[width=0.49\textwidth]{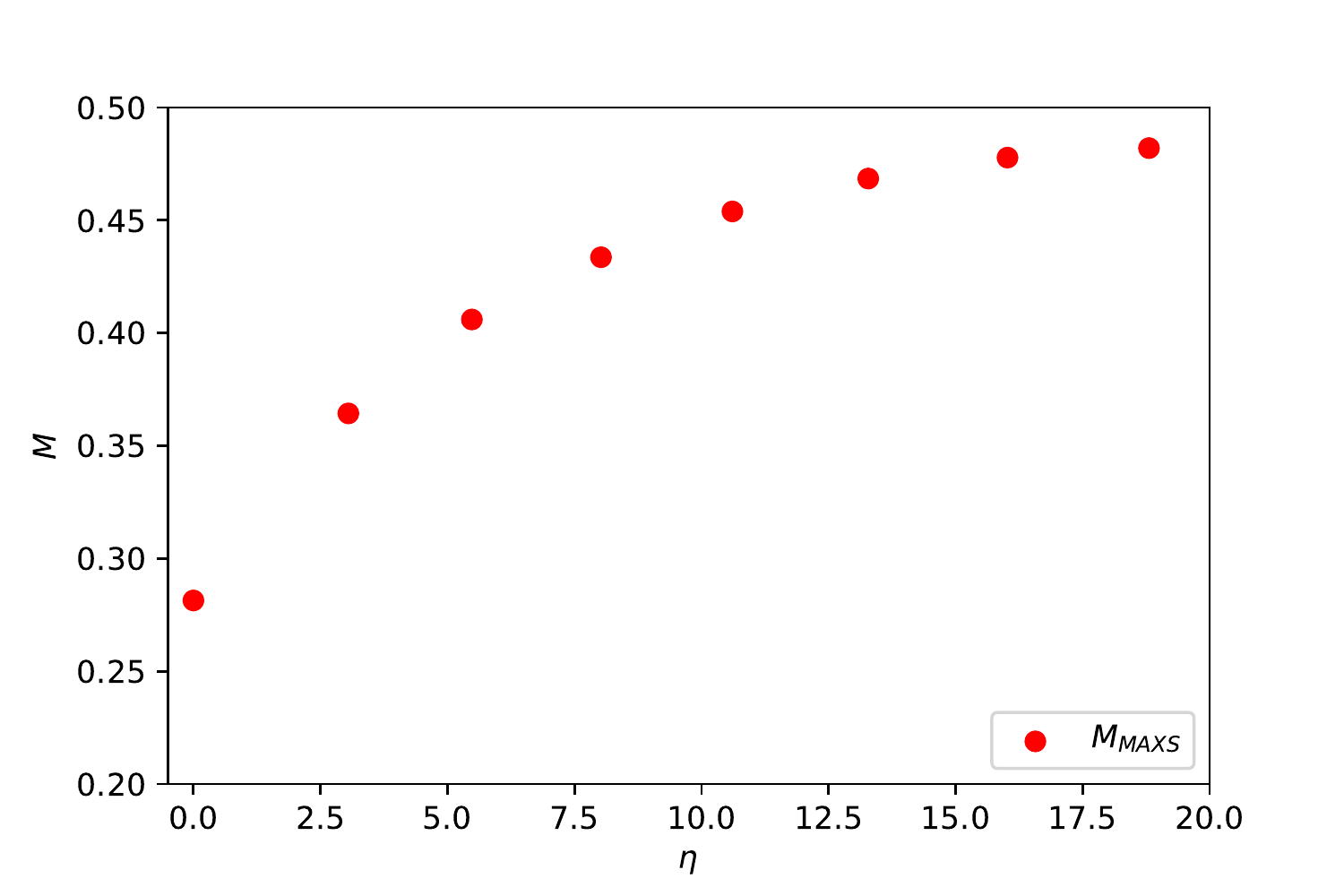}
\caption{Mass of the density peaks for the asymptotic self-similar density profile normalized to $\rho(0)=1$}
 \label{fig:mass-peaks-asymptotic}
 \end{figure}

At large radius, we can neglect the right-hand side in Eq.(\ref{eq:asymptotic}).
This gives a homogeneous equation that can be solved as
\ba
&& \eta \to\infty : \;\;\; \rho \simeq \frac{3b}{2\eta^2} \cos^2 \left[ \frac{a (\eta-\eta_0)}{\sqrt{2}} \right] ,
\nonumber \\
&& M \simeq M_0 + b \left[ \frac{\eta-\eta_0}{2} + \frac{\sin[\sqrt{2} a (\eta-\eta_0)]}{2\sqrt{2}a} \right] ,
\ea
where the parameters $M_0,\eta_0,a$ and $b$ are undetermined (one needs to solve the complete
equation (\ref{eq:asymptotic}) to derive their value).
We can check that the density is always positive but now vanishes on an almost periodic
set of radii.
Thus, the oscillations of the asymptotic profile have equal length in the radius $\eta$,
while the linear-profile oscillations had equal lengths in $\eta^2$, as seen in
Eq.(\ref {eq:delta-L4-large-eta}). The nonlinear effects both move the density peaks
towards the center and change their scaling with distance.
The enveloppe of the density oscillations decreases like $1/\eta^2$, while the soliton density
shows an exponential falloff.
This also implies that the mass grows linearly with the radius, so that each peak (or more precisely
each shell in the 3D space) contains the same mass.
Thus, the outer shells are not as negligible as appears in the density plots.
This is also shown in Fig.~\ref{fig:mass-peaks-asymptotic}, where we plot the mass associated
with the first few density peaks (i.e. the mass in each spherical shell delimited by minima of the
density).

\section{CDM comparison, semiclassical limit and conclusion}
\label{conc}

As we recalled in Eqs.(\ref{eq:CDM-def})-(\ref{eq:CDM-nonlinear}), 
self-similar solutions for collisionless matter in a perturbed Einstein-de Sitter universe
have been obtained by \cite{Fillmore:1984wk}. Using a different method, these
results were recovered and extended to collisional gas by \cite{Bertschinger:1985pd}.
These solutions describe the gravitational collapse of spherical overdense regions
or the expansion of voids. Thus, starting at early times with a small linear perturbation
$\delta_L(r,t_i)$, which follows the power-law profile (\ref{eq:CDM-def}), 
the density contrast grows as $t^{2/3}$,
according to the linear growing mode, until it reaches the nonlinear regime.
Then, nonlinear effects modify the shape of the profile in the inner regions.
At small radii it takes again a power-law form, as in Eq.(\ref{eq:CDM-nonlinear}), 
but with a different exponent that depends on the slope of the linear seed.
Thus, one obtains a family of solutions, characterized by the slope $\gamma$ 
of the density perturbation
at large distance. For the collisional case, the profile also depends on the adiabatic index
$\gamma_{\rm ad}$.
In the case of overdense regions, these self-similar solutions display a gravitational
instability and increasingly distant shells collapse. They typically stabilize at a fixed
fraction of their turnaround radius, as gravity is balanced either by the radial velocity
dispersion or by the thermal pressure (in the collisional case).
This leads to a virial equilibrium in the inner nonlinear core, with a mass and a radius
 that grow with time, both in physical and comoving coordinates.

The self-similar solutions we have obtained in this paper for FDM are very different.
They do not follow power-law shapes and their amplitude does not grow with time.
Thus, the density contrast at the center, $\delta(0)$, remains constant.
This implies that the gravitational instability is balanced by the quantum pressure
and there is no transition from the linear to the nonlinear regime for the profile as time increases.
The profile remains linear on all scales and at all times or remains nonlinear in the center.

In contrast with the CDM case, the large-distance and linear-theory limits
no longer coincide. Although at large radii the density perturbation becomes small and
a linear treatment is valid, the profile does not converge to the linear-theory profile.
This is because at large distance we have nonzero contributions from the three linear
modes that are well behaved at infinity, while the linear theory selects only one of them, which
satisfies in addition the boundary condition at the center.

The constant amplitude with time also means that these solutions do not exhibit a gravitational
collapse. In fact, both for underdense and overdense central regions, the mass contained
in the central peak or ``void'' decreases with time as $M \propto t^{-1/2}$.
The associated radius increases as $t^{1/2}$ in physical coordinates but decreases as
$t^{-1/6}$ in comoving coordinates.
Thus, instead of accreting mass, the central region continuously expels matter.
In the nonlinear regime, this occurs through well-separated clumps that travel outward
in physical coordinates, in a fashion similar to the expulsion of matter in gravitational
cooling.
However, in these self-similar solutions matter does not remain trapped inside each
clump and moves outward faster, leaking from one clump to the next one
through a low-density but high-velocity narrow region.
At large radii, in the linear regime, we have acoustic waves around the cosmological
background as the central overdensity is screened as for a compensated profile,
so that the gravitational force is negligible as compared with kinetic and quantum-pressure terms.

The characteristic exponents $M \propto t^{-1/2}$, $r \propto t^{1/2}$,
are universal, in contrast with the continuous range of exponents
obtained in the CDM case, which depend on the slope $\gamma$ of the linear seed.
As explained in Sec.~\ref{sec:CDM-linear}, this is due to the new force associated with
the quantum pressure, in addition to gravity.
This new term in the equations of motion is only compatible with the exponent $\gamma=-4$.
However, this value is beyond the allowed range  Eq. (\ref{eq:CDM-def}) for the standard 
CDM self-similar solutions. In fact, the self-similar FDM solutions studied in this paper
are instead associated with a non-standard CDM solution, which would correspond to a
decaying mode in the linear regime and as such is unphysical.
However, the separation between growing and decaying modes disappears in the FDM regime,
where the quantum pressure turns both modes into acoustic oscillations of constant
amplitude.
Thus, the FDM case shows strong qualitative differences with the CDM case.
Because of self-similarity, these differences remain important in the solutions studied
in this paper at all times and scales. They do not disappear either in the limit $\epsilon\to 0$
because $\epsilon$ is fully absorbed by the change to the self-similar
coordinate $\eta$ in Eq.(\ref{eq:eta-def-x}).

As explained above, the profile at large distance, although within the linear regime of
small perturbations, differs from the linear-theory profile by two additional linear modes,
with coefficients that depend on the central density.
Thus, in contrast with the CDM case, we have a strong coupling from the small inner radii
onto the large outer radii.
Indeed, in the CDM non-collisional case \cite{Fillmore:1984wk,Bertschinger:1985pd}, outer shells are decoupled and follow the
spherical collapse, which only depends on the inner mass from Gauss theorem
and not on the density profile.
This means that they collapse as in free fall (with an initial outward velocity close to the
Hubble flow) until shell crossing occurs at nonlinear radii (which implies the mass within each
shell is no longer constant and the dynamics become more complex).
Similarly, in the collisional collapse of a polytropic gas, the pressure at large radii beyond the
shock (located around the virial radius) is zero (because one starts from cold initial conditions),
so that free-fall spherical collapse also applies. The pressure becomes nonzero at the shock,
associated with a jump of the temperature and of the entropy, and balances gravity at smaller
radii \cite{Bertschinger:1985pd,Teyssier1997}.
In contrast, for the self-similar solutions of FDM, outer shells do not follow a spherical free fall.
This is because the effective ``pressure'', associated with the quantum pressure, is not zero.
Instead, together with kinetic terms, it dominates over gravity. This leads to acoustic-like
oscillations, or waves, that transport information from small to large scales and build this
coupling between small and large scales.

As the Schr\"odinger equation (\ref{eq:psi-eq-comoving}) was also proposed 
\citep{Widrow1993,Uhlemann2014,Mocz2018,Garny2020}
as an alternative
to N-body simulations to compute the evolution of CDM in the semiclassical limit,
$\epsilon\to 0$, one can be surprised by this large difference between the self-similar solutions.
Indeed, in the semiclassical limit one expects to recover the CDM dynamics.
The point is that this is a weak limit that is not so straightforward.
As shown by the scalings (\ref{eq:self-r-x}) and (\ref{eq:eta-def-x}),
what happens is that these FDM self-similar solutions ``disappear'' in the semiclassical limit,
as their size and their mass decrease as $\epsilon^{1/2}$ and $\epsilon^{3/2}$ for $\epsilon\to 0$.
Thus, in the semiclassical limit, they become confined to an increasingly small radius.
This counterbalances the $\epsilon^2$ factor in front of the Laplacian in the
Schr\"odinger equation (\ref{eq:psi-eq-comoving}), so that for any finite $\epsilon$
we have a self-similar solution where the quantum pressure remains able to balance gravity
near the center. However, from a macroscopic point of view, this configuration becomes
irrelevant as it becomes infinitesimal.
The standard CDM self-similar solutions are only exactly recovered at $\epsilon=0$, or as approximate solutions
at small $\epsilon$ with a small breaking of the self-similarity.
In other words, the standard CDM self-similar solutions are not the limit at $\epsilon \to 0$
of the FDM self-similar solutions.
This also shows that the semiclassical limit needs to be considered with care.
If in the limit $\epsilon \to 0$ gradients become steep enough, they can sustain
dynamics quite different from the CDM Vlasov case and the limit can become nontrivial.

\acknowledgments

R.G.G. was supported by the CEA NUMERICS program, which has received funding from the European Union's Horizon 2020 research and innovation program under the Marie Sklodowska-Curie grant agreement No 800945.

This project has received funding /support from the European Union's Horizon 2020 research and innovation programme under the Marie Sklodowska -Curie grant agreement No 860881-HIDDeN.

This work was
made possible by with the support of the Institut Pascal at University Paris-Saclay during the 
Paris-Saclay Astroparticle Symposium 2021, with the support of the P2IO Laboratory of Excellence (program ``Investissements d'avenir'' ANR-11-IDEX-0003-01 Paris-Saclay and
ANR-10-LABX-0038), the P2I axis of the Graduate
School Physics of University Paris-Saclay, as well as IJCLab, CEA, IPhT, APPEC, the IN2P3 master projet UCMN and EuCAPT ANR-11-IDEX-0003-01 Paris-Saclay and ANR-10-LABX-0038.

\bibliography{ref}

\end{document}